\documentclass[aps,pra,twocolumn,superscriptaddress,floatfix,showkeys,amsmath,amssymb,longbibliography,colorlinks,urlcolor=blue,citecolor=blue,linkcolor=blue]{revtex4-2}

\usepackage{bbold}          
\usepackage{graphicx}        
\usepackage{braket}          
\usepackage{cancel}          
\usepackage{dcolumn}         
\usepackage{bm}              
\usepackage{tikz}            
\usetikzlibrary{arrows,shapes,positioning,shadows,svg.path,backgrounds,fit,calc}
\usepackage[normalem]{ulem}  
\usepackage{comment}         
\usepackage{subfig}          
\usepackage{orcidlink}       

\usepackage[font=small,labelfont=bf,format=plain,justification=centerlast,labelsep=period]{caption}

\begin{document}

\title{Ergotropy from energetic coherence and the third law of thermodynamics}

\author{O. S. Y\'a\~nez-Soria\,\orcidlink{0009-0009-9399-7156}}
\affiliation{Centro de F\'isica Aplicada y Tecnolog\'ia Avanzada, Universidad Nacional Aut\'onoma de M\'exico, Boulevard Juriquilla 3001, Quer\'etaro 76230, Mexico}

\author{G. Guarnieri\,\orcidlink{0000-0002-4270-3738}} 
\affiliation{Dipartimento di Fisica ``A. Volta", Universit\`a di Pavia, via Bassi 6, 27100 Pavia (Italy)}

\author{B. \c{C}akmak\,\orcidlink{0000-0002-6124-3925}} 
\affiliation{Department of Physics, Farmingdale State College SUNY, Farmingdale, NY 11735, USA}

\author{R. Rom\'an-Ancheyta\,\orcidlink{0000-0001-6718-8587}}
\email{ancheyta@fata.unam.mx}
\affiliation{Centro de F\'isica Aplicada y Tecnolog\'ia Avanzada, Universidad Nacional Aut\'onoma de M\'exico, Boulevard Juriquilla 3001, Quer\'etaro 76230, Mexico}

\date{\today}

\begin{abstract}
We study the ergotropy and the thermodynamic potentials of a qubit battery charged by repeated interactions with clusters of thermal ancillas, qubits or oscillators, through a composite system-bath coupling that generates steady-state coherence in the energy basis. A single parameter encoding the bath quantum statistics yields the coherence and ergotropy in closed form for any cluster size. For large clusters, the ergotropy becomes independent of the bath statistics and, at optimal couplings, reaches a universal fraction of the equilibrium internal energy of the qubit at the bath temperature. Since the charged state has no population inversion, all extractable work comes from coherence. The ergotropy is flat at low temperatures. We trace this plateau to the third law, and show that the ergotropy is proportional to the equilibrium enthalpy of the qubit, whose slope vanishes with the heat capacity. For a battery composed of multiple qubits, the coherence grows exponentially with the number of qubits while the ergotropy grows linearly, both remaining a nearly constant fraction of their respective maxima. The charged state is a non-equilibrium steady state with finite entropy at zero temperature, sustained by the work of switching the collisional interaction on and off. The cost is nonzero whenever the parallel coupling is nonzero, and, unlike the ergotropy, it depends on the bath statistics.
\end{abstract}

\maketitle

\section{Introduction}\label{sec:intro}

Quantum batteries are devices in which energy is stored in, and extracted from, a quantum system, with the potential to exploit genuinely quantum resources such as coherence and entanglement to outperform classical storage in energy density or charging speed \cite{RMP_Campaioli_2024,Razzoli_2025,Myers_2022,Review-en-Baterias-convencionales-2015,Quantum-vs-classical-many-body-QB-2019,Colloquium-QB-2024,rinaldi2025reliable,rinaldi2026maximum}. Since any realistic battery interacts with its surroundings, it must be treated as an open system. The two-level system, or qubit, is the minimal model for such an analysis and the natural building block for larger architectures \cite{Many-body-QB-2022}. Its weak-coupling dynamics can be described by a Markovian master equation obtained, for instance, from repeated system-bath interactions (also known as the quantum collision model)~\cite{Guarnieri-memoryless-2020,Giovannetti-TLScase-2019,Carrega-Dissipative-TLScase-2020,Geraldine-Haack-Quantum-Advantage-Charging-2023}. The resulting equation, typically in Lindblad form, accommodates dissipation \cite{F.Barra-Dissipative-Charging-QB-2019} and decoherence \cite{QRE-TuncerOzdemir2020} at finite temperature and connects the problem of energy storage to quantum thermodynamics at large \cite{Campbell-Roadmap-Q-Thermodynamics-2025,Vinjanampathy-Quantum-thermodynamics-2016}.

One figure of merit of a battery is the work it can deliver. In macroscopic thermodynamics, the maximum extractable work is given by the free energy difference $\Delta F$ between a non-equilibrium initial state and a final Gibbs state. However, for a finite quantum system, the corresponding quantity is the ergotropy $\mathcal{W}$, defined as the maximum energy extractable by means of unitary operations, which leave the entropy unchanged~\cite{Allahverdyan-2004}. The remaining internal energy is passive and inaccessible without additional resources~\cite{Kamin-Alan-Santos-Exergy-Passive-energy-2021}. Ergotropy can be split into an incoherent part, associated with population inversion, and a coherent part, associated with off-diagonal elements of the density matrix in the energy eigenbasis~\cite{Guarnieri-Coherence-&-Ergotropy-2020,Coherence-Cakmak-2020,Akira-Deffner-Q-&-classical-ergotropy-from-entropy-2021,Smith2022,Campisi_2026}. In multipartite systems, correlations between the constituents provide a further contribution, allowing work to be extracted globally that is inaccessible from the parts separately~\cite{Touil_2022,Salvia_2022,Mula_2023,Biswas_2025,Francica_2025,Dara_2025}. 
A thermal Gibbs state is diagonal and passive, and therefore carries none of these contributions. A thermal environment that merely relaxes the battery to such a state cannot charge it. It is therefore of interest to identify system-bath couplings through which a thermal environment, taking the role of the charger, may still charge the battery.

The class of so-called `composite system-bath interactions' introduced in Ref.~\cite{Guanieri-composite-system-bath-interaction-2018,purkayastha2020tunable}, containing both a component that commutes with the system Hamiltonian and one that does not, can generate energetic coherence (coherence between states of different energy~\cite{Manzano-energetic-coherence}) in the steady state of a system coupled to a thermal bath. In the collision-model realization of this mechanism \cite{R-Ancheyta-Enhanced-2021}, the target battery system interacts repeatedly with thermal ancillary systems (bath elements acting as chargers) and steady-state coherence is enhanced when the latter are made of clusters that collectively `composite-interact' with the system. The same reference also showed that maintaining such steady-state coherence requires a non-zero power input.

In this work, we use that mechanism to charge a qubit battery and ask how much work it stores, at what cost, and what its thermodynamic behavior is.
The charging process is mediated by a thermal bath made of clusters of arbitrary size $N$ of either qubits or harmonic oscillators, simultaneously treated through a statistical parameter $\lambda=\pm1$, sequentially interacting with the battery. We obtain the steady-state ergotropy in closed form, optimized over the coupling constants, and demonstrate that it is entirely of coherent origin. We then examine its temperature dependence, in particular showing that it always reaches a plateau at low temperatures. Turning our attention to this behavior, we conclude that the plateau follows from the third law of thermodynamics (Nernst’s postulate)~\cite{Callen1985,StrasbergBook}, since the optimized ergotropy is proportional to the equilibrium enthalpy of the qubit, whose low-temperature slope vanishes with the heat capacity. In contrast, we show that the free energy difference has a finite slope due to the residual entropy of the non-thermal steady state. Finally, we compute the switching work that the collision model requires to charge and hold the battery, and find that it is nonzero whenever the parallel component of the composite-interaction is nonzero, and that, unlike the ergotropy, it depends on the bath statistics.

The paper is organized as follows. Section~\ref{sec:Steady state for a qubit battery} introduces the model and the steady state. Section~\ref{sec:Ergotropy for a qubit} computes and optimizes the ergotropy, analyzes its temperature dependence, and derives the bounds that internal energy and coherence impose on it. Section~\ref{sec:thermo} contains the thermodynamic analysis, namely the entropy and free energy of the charged state, the origin of the low-temperature behavior, and the thermodynamic cost of charging. Sections~\ref{subsec:Battery-Scaling} and \ref{sec:conclusions} cover the battery scaling with the number of qubits and our conclusions. Derivations are collected in the appendices.

\section{Qubit battery model}\label{sec:Steady state for a qubit battery}

We consider a two-level battery with Hamiltonian $\mathcal{H}_{S}^{}=\hbar\omega_{0}^{}\sigma_z^{}/2$. The bath consists of a stream of clusters, each composed of $N$ mutually non-interacting ancillas. The battery qubit sequentially interacts with these clusters for a short time $\tau$, in the spirit of a quantum collision model or repeated interaction scheme. This model provides one of the most transparent and straightforward microscopic pictures in which the individual characteristics of the system and bath can be specified independently~\cite{Ciccarello-Q-Collision-Model-Review-2022}. We consider clusters, of sizes $N$ ranging from $1$ to $N\gg1$, of two types of ancillas: qubits and quantum harmonic oscillators. Each ancilla is prepared in the Gibbs state $\rho_{_{\rm B}}^{}=\exp(-\beta\,\mathcal{H}_{\!B}^{})/{Z}$ at inverse temperature $\beta\equiv(k_{\!B}^{}T)^{-1}$, where $Z={\rm tr}\{\exp(-\beta\,\mathcal{H}_{B}^{})\}$ is the partition function and $\mathcal{H}_{\!B}^{}$ denotes the free Hamiltonian of the corresponding type of ancilla (Appendix~\ref{Apx:Reservoir}). The total Hamiltonian during a collision is $\mathcal{H}=\mathcal{H}_{S}^{}+\mathcal{H}_{B}^{}+\mathcal{H}_{I}^{}.$
The interaction is rescaled by the collision time, $\mathcal{H}_{I}^{}=\mathcal{V}_{I}^{}/\sqrt{\mathcal{\tau}}$, which is the standard choice that yields a well-defined Markovian limit as $\tau\to0$ ~\cite{Ciccarello-Q-Collision-Model-Review-2022}. The operator $\mathcal{V}_{I}^{}$ which we considered belongs to the class of `composite-interactions' characterized in Refs.~\cite{Guanieri-composite-system-bath-interaction-2018, purkayastha2020tunable, R-Ancheyta-Enhanced-2021}
 \begin{equation}\label{eq:Vi-General-Interaction-Hamiltonian}
\mathcal{V}_{I}^{}=f_1^{}\sigma_z\otimes(B_{\lambda}^{}+B^\dagger_{\lambda})+f_2^{}(\sigma_+\otimes B_{\lambda}^{}+\sigma_-\otimes B^\dagger_{\lambda}),
\end{equation}
with $f_1^{}$ and $f_2^{}$ real coupling constants. Within the rotating-wave approximation, the first term is the component \emph{parallel} to $\mathcal{H}_{S}^{}$, because it commutes with it, and the second is the \emph{orthogonal} component ~\cite{Guanieri-composite-system-bath-interaction-2018}. 
The parallel term alone ($f_2^{}=0$) dephases the battery and creates coherence in the ancilla; the orthogonal term alone ($f_1^{}=0$) exchanges excitations and can transfer to the qubit any coherence the ancilla already carries. Because the ancilla is in an incoherent thermal state, neither terms by itself produce coherence in the steady state when acting separately. However, when simultaneously activated ($f_1^{}\cdot f_2^{}\neq0$), a non-trivial interplay takes place and the coherence created in the ancilla by the parallel term is fed back into the battery by the orthogonal one, thus leading to the formation of steady-state coherence in the battery energy eigenbasis. We will refer to this as steady-state coherence (SSC).

Along the lines of Ref.~\cite{R-Ancheyta-Enhanced-2021}, we consider the bath operator $B_{\lambda}^{}\equiv\sum_{i=1}^{N}b^{{(i)}}_{\lambda}$ to be a collective operator for the $N$ ancillas of a cluster (Appendix~\ref{Apx-sub:Generic-Baths}). We introduce a parameter $\lambda$ that encodes the statistics of the ancillas through the relation $b_\lambda^{}b_\lambda^\dagger-\lambda\, b_\lambda^\dagger b_\lambda^{}=1$, following the parametrization used in \cite{Moroni-Q-thermal-machine-rectifier-2025,Palafox-Heat-2022} to study heat rectification via quantum statistical and coherent asymmetries. In particular, for $\lambda=1$, $b_{\lambda}^{}$ and $b_{\lambda}^{\dagger}$ are the ladder operators of a harmonic oscillator and the thermal occupation is the Bose-Einstein one ($n_{{\rm B}}^{}$); for $\lambda=-1$ they are the lowering ($\sigma_-^{B}$) and raising ($\sigma_+^{B}$) operators of a bath qubit and the corresponding occupation number is the Fermi-Dirac one $n_{\rm F}^{}$. Both cases are compactly captured by 
\begin{equation}\label{eq:DistributionFunction}
    n(\lambda)=\frac{1}{\exp{\left(\hbar\omega_{B}^{}/k_{\!B}^{}T\right)}-\lambda}.
\end{equation}
We restrict ourselves to the resonant case $\omega_{0}^{}=\omega_{B}^{}\equiv\omega$. Writing Eq.~\eqref{eq:Vi-General-Interaction-Hamiltonian} as a bilinear system–bath interaction $\mathcal{V}_{I}^{}=s^\dagger B_{\lambda}^{}+sB_{\lambda}^{\dagger}$ with $s\equiv f_1^{}\sigma_z^{}+f_2^{}\,\sigma_-^{}$, the reduced dynamics of the qubit battery is given by the master equation~\cite{R-Ancheyta-Enhanced-2021,roman2019spectral}:
\begin{multline}\label{eq:GenericMasterEquation}
 \frac{d\rho}{dt}=-\frac{i\omega}{2}[\sigma_z^{},\rho]+\langle B_{\lambda}^{}B_{\lambda}^{\dagger}\rangle\,\mathcal{L}[f_1^{}\sigma_z^{}+f_2^{}\sigma_-^{}]\rho
\\
+\langle B_{\lambda}^{\dagger}B_{\lambda}^{}\rangle\,\mathcal{L}[f_1^{}\sigma_z^{}+f_2^{}\sigma_+^{}]\rho.
\end{multline}
Here $\mathcal{L}[x]\rho\equiv x\rho x^\dagger-\frac{1}{2}(x^\dagger x\rho+\rho x^\dagger x)$ is the Lindblad superoperator and the corresponding expectation values are computed with respect to the thermal state of the bath cluster. Note that the last two terms in Eq.~(\ref{eq:GenericMasterEquation}) do not describe pure absorption and emission channels. Their arguments mix parallel and orthogonal components, and it is this mixing that generates SSC.

The state of the qubit can be described in terms of the Bloch vector $\langle\vec{\sigma}\rangle=(\langle\sigma_x\rangle,\langle\sigma_y\rangle,\langle\sigma_z\rangle)$. Its equations of motion and steady-state solution follow from Eq.~\eqref{eq:GenericMasterEquation} and are given in Appendix~\ref{Apx:Bloch Equations}. To quantify the generated coherence we use the $l_1$-norm \cite{Coherence-l1norm-2014}, $\mathcal{C}(t)=\sum_{i\neq j}|\rho_{{i,j}}^{}(t)|$, which for a qubit reduces to $\mathcal{C}(t)=|\langle\sigma_x(t)\rangle + i\langle\sigma_y(t)\rangle|$. Inserting the steady-state values $\langle\sigma_x^{\lambda}\rangle_{\rm _{SS}}$ and $\langle\sigma_y^{\lambda}\rangle_{\rm _{SS}}$ from Eqs.~\eqref{eq:GenericSigmaX} and \eqref{eq:GenericSigmaY}, the SSC for arbitrary cluster size and bath elements is
\begin{equation}\label{eq:Generic_SS_Coherence}
        \mathcal{C}_{{\rm SS}}^{}(\lambda)=f_1^{}f_2^{}\frac{r(\lambda)}{s(\lambda)+\omega^2/N},
\end{equation}
where $s(\lambda)\!\equiv\!(2f_1^2+f_2^2/2)(f_1^2+f_2^2/2)N[1+(\lambda+1)n(\lambda)]^{2}$,
$r(\lambda)\!\equiv\![1\!+\!(\lambda\!-\!1)n(\lambda)]|N(2f_1^2\!+\!f_2^2/2)\left[1\!+\!(\lambda\!+\!1)n(\lambda)\right]\!+\!i\omega|$
are two temperature dependent functions. 

In the large-cluster limit $N\gg1$, $\langle\sigma_y^\lambda\rangle_{_{\rm SS}}$ is of order $1/N$ and may be neglected. The remaining components of the Bloch vector then take the simple form
$\langle\sigma_x^\lambda\rangle_{_{\rm SS}}^{^{\!N\gg1}}= \mathcal{C}_0^{} \tanh\big(\hbar\omega/{2k_{\!B}^{}T}\big)$ and $\langle\sigma_z^\lambda\rangle_{_{\rm SS}}^{^{\!N\gg1}}=(\mathcal{C}_0{f_1^{}}/{f_2}-1) \tanh({\hbar\omega}/{2k_{\!B}^{}T})$
where, as in \cite{R-Ancheyta-Enhanced-2021},
\begin{equation}\label{eq:C_zero}
\mathcal{C}_0^{}\equiv\frac{f_1^{}f_2^{}}{f_1^2+f_2^2/2}.
\end{equation}
For $f_1^{}=0$ only the dissipative channel in Eq.~(\ref{eq:GenericMasterEquation}) is active and $\rho_{_{\rm SS}}^{}$ reduces to a Gibbs state at the bath temperature. For $f_1^{}\cdot f_2^{}\neq0$, $\rho_{_{\rm SS}}^{}$ has off-diagonal elements in the energy basis and is a non-equilibrium steady state.  Note that in this limit the steady state loses all memory of the nature of the ancillas and is the same for bath oscillators and qubits, which can be understood through the Holstein-Primakoff mapping of collective spin operators onto bosonic ones \cite{R-Ancheyta-Enhanced-2021}. Accordingly, Eq.~\eqref{eq:Generic_SS_Coherence} reduces to
\begin{equation}\label{eq:Coherence_SS_N-limit}
    \mathcal{C}_{\rm{SS}}^{^{N\gg1}}= \mathcal{C}_0 \tanh\big({\hbar\omega}/{2k_{\!B}^{}T}\big).
\end{equation}
Here $\mathcal{C}_0$ is the zero-temperature steady-state coherence in the large-cluster limit. It is nonzero if and only if $f_1^{}\cdot f_2^{}\neq0$.

\section{Ergotropy for a qubit battery}\label{sec:Ergotropy for a qubit}

The ergotropy $\mathcal{W}$ of a state $\rho$ is the maximum work that can be extracted from it by unitary operations \cite{Allahverdyan-2004}. It equals the energy difference between $\rho$ and its passive state $\sigma_{\!\rho}^{}=\sum_i^{} r_{i}^{}|E_{i}^{}\rangle\langle E_{i}^{}|$, where $r_{i}^{}$ are the eigenvalues of $\rho=\sum_i r_{i}^{}|r_{i}^{}\rangle\langle r_{i}^{}|$ and $|E_{i}^{}\rangle$ the eigenvectors of the system Hamiltonian $\mathcal{H}=\sum_j\varepsilon_{j}^{}|E_{j}^{}\rangle\langle E_{j}^{}|$, both ordered so that the largest population sits on the lowest energy level, i.e., $r_{1}^{}\geq r_{2}^{}\geq\ldots$ and $\varepsilon_{1}^{}< \varepsilon_{2}^{}<\ldots$. With $\langle\mathcal{H}(\rho)\rangle\equiv{\rm tr}\{\rho\,\mathcal{H}\}$, $\mathcal{W}(\rho)=\langle\mathcal{H}(\rho)\rangle-\langle\mathcal{H}(\sigma_{\!\rho})\rangle$~\cite{F.Barra-Dissipative-Charging-QB-2019}, or, equivalently, for a $d$-dimensional system \cite{Allahverdyan-2004}, $\mathcal{W}(\rho)=\sum_{i,j}^{}r_{i}^{}\varepsilon_{j}^{}\left(|\langle r_{i}^{}|\varepsilon_{j}^{}\rangle|^{2}-\delta_{{ij}^{}}\right)$, with $\{r_i^{},\varepsilon_j^{}\}$ ordered as above. For our qubit battery with $\mathcal{H}_{S}^{}=\hbar\omega\sigma_z^{}/2$, the eigenvalues of $\rho$ are $r_{1,2}^{}=(1\pm\|\langle\vec{\sigma}\rangle\|)/2$, which yields
\begin{equation}\label{eq:Ergotropy_BlochVector}
   \mathcal{W}={\hbar\omega}\left(\langle\sigma_z^{}\rangle+\| \langle\vec{\sigma}\rangle\|\right)/2,
\end{equation}
where $\|\langle\vec{\sigma}\rangle\|$ is the length of the Bloch vector. For a Gibbs state, which is the steady state of Eq.~(\ref{eq:GenericMasterEquation}) when $f_1^{}=0$, one has $\langle\sigma_x^{}\rangle=\langle\sigma_y^{}\rangle=0$, $\langle\sigma_z^{}\rangle=-\tanh(\hbar\omega/2k_{\!B}^{}T)$, and Eq.~\eqref{eq:Ergotropy_BlochVector} reduces to $\mathcal{W}_{\rm inc}={\hbar\omega}\left(\langle\sigma_z^{}\rangle+|\langle\sigma_z^{}\rangle|\right)/2$, which vanishes for $\langle\sigma_z^{}\rangle<0$, as it must for a passive state \cite{Allahverdyan-2004}. $\mathcal{W}_{\rm inc}$ holds for any incoherent state $\rho=p_e^{}|e\rangle\langle e|+p_g^{}|g\rangle\langle g|$, for which $\langle\sigma_z\rangle=p_e^{}-p_g^{}$: the ergotropy is zero for $p_e^{}\leq p_g^{}$ and equals $\hbar\omega\langle\sigma_z\rangle$ for a population inversion $p_e^{}>p_g^{}$. As an example, the inverted Gibbs state $\exp(+\beta\,\mathcal{H}_S^{})/{\rm tr}\{e^{+\beta H_S}\}$ used in \cite{F.Barra-Dissipative-Charging-QB-2019} stores $\hbar\omega\tanh(\hbar\omega/2k_{\!B}^{}T)$. This is an incoherent contribution to the ergotropy.

The coherent contribution enters through $\langle\sigma_x^{}\rangle$ and $\langle\sigma_y^{}\rangle$, which are also the components that determine the $l_1$-norm of coherence. Writing the internal energy as $U=(\hbar\omega/2)\langle\sigma_z\rangle$ and defining an energy-scaled coherence $C\equiv(\hbar\omega/2)\,\mathcal{C}$, for the steady state of the model described in Sec.~\ref{sec:Steady state for a qubit battery}, Eq.~\eqref{eq:Ergotropy_BlochVector} can be recast as follows~\cite{Choquehuanca-2024}
\begin{equation}\label{eq:Ergotropy(lambda)} 
\mathcal{W}_{{\rm SS}}^{}(\lambda)=U_{{\rm SS}}^{}(\lambda)+\sqrt{\,U_{{\rm SS}}^{{\,2}}(\lambda)+C_{{\rm SS}}^{{\,2}}(\lambda)\,},
\end{equation}
where $U_{{\rm SS}}^{}(\lambda)=(\hbar\omega/2)\,\langle\sigma_z^{\lambda}\rangle_{\!_\text{SS}}$ can be obtained from Eq.~\eqref{eq:GenericSigmaZ} and $C_{{\rm SS}}^{}(\lambda)=({\hbar\omega}/{2})\mathcal{C}_{{\rm SS}}^{}(\lambda)$ from Eq.~(\ref{eq:Generic_SS_Coherence}). Equation~\eqref{eq:Ergotropy(lambda)} is the ergotropy of our qubit battery for any cluster size $N$ and either bath statistics. Bounds that follow from it, and their representation on the Bloch sphere, will be discussed in Sec.~\ref{subsec:Bounds}.

\subsection{Large-cluster limit and optimal couplings}\label{subsec:optimum}

\begin{figure}[t]
    \centering
  \begin{tikzpicture}
        \node (img){\includegraphics[width=0.48\textwidth]{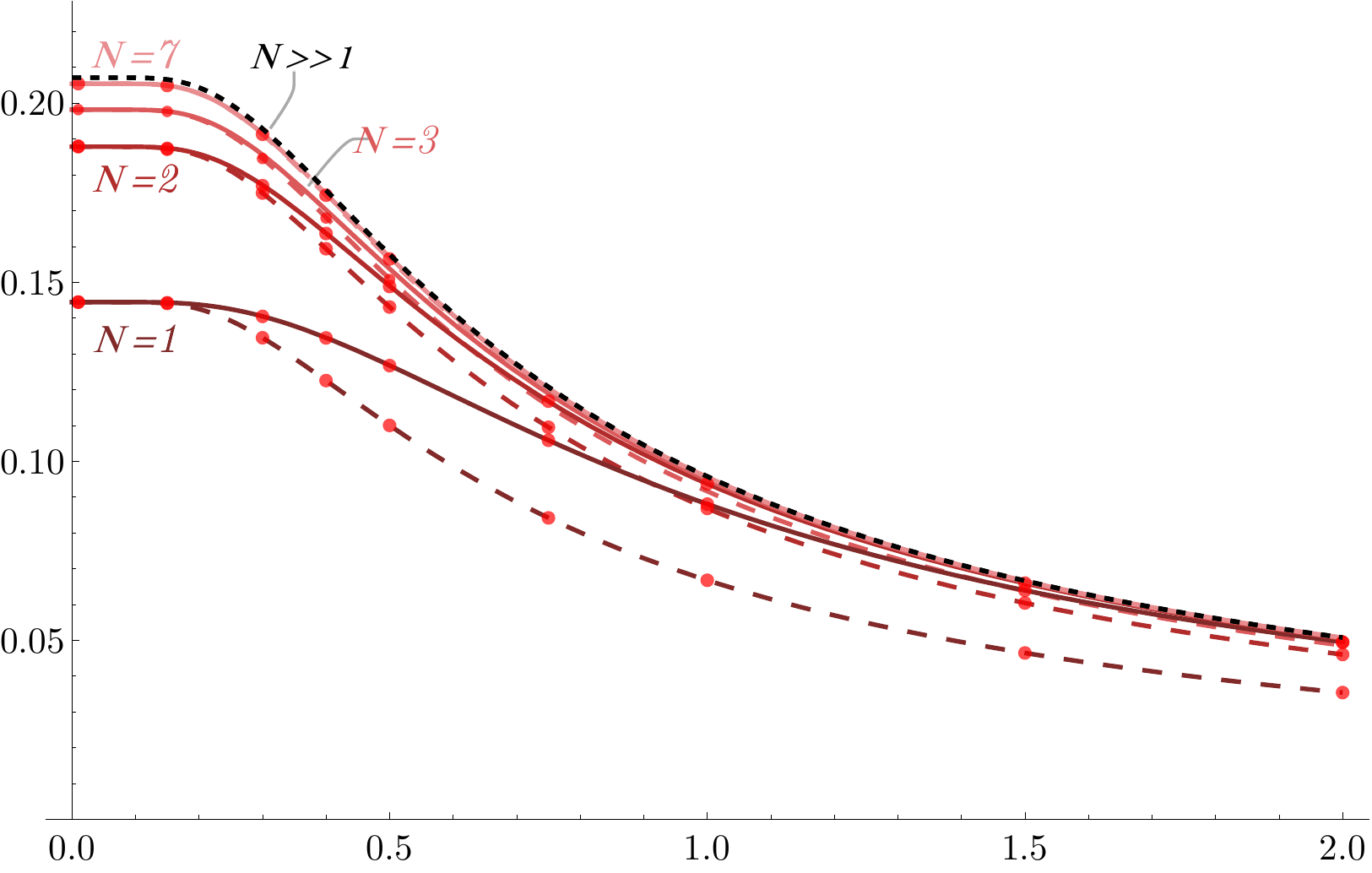}};
        \node[above=of img, node distance=0cm, xshift=0cm, yshift=-1.5cm]
        {Ergotropy ($\mathcal{W}_{{\rm SS}^{}}(\lambda)/\hbar\omega$)};
        \node[above=of img, node distance=0cm, xshift=0cm, yshift=-7.2cm]
        {$k_{\!B}^{}T/\hbar\omega$};
    \end{tikzpicture}
     \caption{Steady-state ergotropy [Eq.~(\ref{eq:Ergotropy(lambda)})] of the qubit battery as a function of the bath temperature for clusters of $N$ qubits (dashed lines) and $N$ oscillators (solid lines), $N=1,2,3,7$, with the couplings fixed by Eq.~(\ref{eq:OptimizationIdentity}). The uppermost short-dashed line is the large-cluster limit, Eq.~(\ref{eq:Optimized_Ergotropy}), which is common to both reservoirs. Note that Eq.~(\ref{eq:OptimizationIdentity}) is the optimum for $N\gg1$; the finite-$N$ curves are not individually optimized.}
    \label{fig:Ergotropy-curves}
\end{figure}

For the ergotropy in the limit $N\gg1$, the internal energy and the energy-scaled coherence are obtained from $\langle\sigma_z\rangle_{_{\rm SS}}^{^{\!N\gg1}}$ and $\mathcal{C}_{\rm{SS}}^{^{N\gg1}}$, both multiplied by $\hbar\omega/2$, reducing Eq.~\eqref{eq:Ergotropy(lambda)} to
\begin{equation}\label{eq:Ergotropy-Co-&-A=f1-f2}
       \mathcal{W}_{{\rm SS}}^{^{N\gg1}}(T)={\hbar\omega}A\,\mathcal{C}_0\tanh\big({\hbar\omega}/{2k_{\!B}^{}T}\big),
\end{equation} 
with the amplitude $A=[(4+f_2^{2}/f_1^{2})^{1/2}-f_2^{}/f_1^{}]/4$. Like the Bloch vector, the ergotropy in this limit is independent of the bath statistics $\lambda$. In Fig.~\ref{fig:Ergotropy-curves} the finite-$N$ curves from Eq.~(\ref{eq:Ergotropy(lambda)}) for both qubits and oscillators baths, converge to Eq.~\eqref{eq:Ergotropy-Co-&-A=f1-f2} as $N$ increases.
Equation~\eqref{eq:Ergotropy-Co-&-A=f1-f2} depends on the couplings only through their ratio. Maximizing $A\,\mathcal{C}_0$ with respect to $f_1^{}/f_2^{}$ gives
\begin{equation}\label{eq:OptimizationIdentity}
   f_1^{}=f_2^{}\big(1+\sqrt{2}\big)^{\!\frac{1}{2}}/\sqrt{2},
\end{equation}
and the optimized ergotropy
\begin{equation}\label{eq:Optimized_Ergotropy}
\mathcal{W}_{{\rm SS}}^{*}(T)=({\hbar\omega/2})\big({\sqrt{2}-1}\big)\tanh\big({\hbar\omega}/{2k_{\!B}^{}T}\big),
\end{equation}
shown as the uppermost dashed curve in Fig.~\ref{fig:Ergotropy-curves}. The same amplitude $(\sqrt{2}-1)/2$ appears in the zero-temperature ergotropy of a qubit charged by a resonant classical driving field~\cite{Morrone-2023}. Here, it emerges instead from the composite interaction with an incoherent bath, and we obtain its full temperature dependence in simple closed form.

At the optimum and at $T=0$, we have $\langle\sigma_x\rangle=(1+\sqrt{2})^{-1/2}\approx0.64$, $\langle\sigma_y\rangle=0$ and $\langle\sigma_z\rangle=1/\sqrt{2}-1\approx-0.29$, so that $U_{{\rm SS}}^{*}/\hbar\omega=1/(2\sqrt{2})-1/2\approx-0.15$, $\mathcal{C}_{{\rm SS}}^{*}\approx0.64$ and $\mathcal{W}_{{\rm SS}}^{*}/\hbar\omega=(\sqrt{2}-1)/2\approx0.207$. The excited-state population is $p_e^{}=(1+\langle\sigma_z\rangle)/2\approx0.35<p_g^{}$, which means that there is no population inversion, and an incoherent state with the same $\langle\sigma_z\rangle$ would have zero ergotropy by $\mathcal{W}_{\rm inc}$. All of the extractable work at the optimum is therefore of coherent origin. The purity of this state is $\mathcal{P}=(1+\|\langle\vec{\sigma}\rangle\|^2)/2=3/4$.

The condition \eqref{eq:OptimizationIdentity} differs from the one that maximizes the coherence itself, $f_1^{}=f_2^{}/\sqrt{2}$~\cite{R-Ancheyta-Enhanced-2021}. The two optima do not coincide because the ergotropy depends on both $C$ and $U$, which are not independent; see Sec.~\ref{subsec:Bounds}. Increasing $f_1^{}/f_2^{}$ beyond the coherence optimum trades some coherence for a less negative internal energy, and Eq.~\eqref{eq:Ergotropy(lambda)} rewards the trade.

\subsection{Temperature dependence and bath statistics}\label{subsec:temperature}

Figure~\ref{fig:Ergotropy-curves} shows that the ergotropy is largest at $T=0$, remains on a plateau over a finite range of low temperatures, and vanishes for $k_{\!B}^{}T/\hbar\omega\gg1$. For finite $N$ the decay with temperature is slower for oscillator baths than for qubit baths. To understand the latter we approximate both thermal occupations by the single high-temperature form~\cite{Palafox-Heat-2022}
\begin{equation}\label{eq:Approx_linear_distributions}
    n(\lambda)\approx\frac{\lambda}{2}\bigg[\bigg(\frac{2k_{\!B}^{}T}{\hbar\omega}\bigg)^{\!\!\lambda}-1\bigg],
\end{equation}
which reproduces the leading terms of $n_{{\rm B}}^{}\approx k_{\!B}^{}T/\hbar\omega-1/2$ for $\lambda=+1$ and of $n_{{\rm F}}^{}\approx1/2-\hbar\omega/(4k_{\!B}^{}T)$ for $\lambda=-1$. Substituting Eq.~\eqref{eq:Approx_linear_distributions} into Eqs.~\eqref{eq:GenericSigmaX}--\eqref{eq:GenericSigmaZ}, the internal energy and the energy-scaled coherence of the steady state at arbitrary $N$ become
\begin{equation}\label{eq:HO_Qubit_Lambda}
U_{{\!\rm SS}}^{}(\lambda)\!\approx\!\frac{\hbar\omega}{2}\! \left[\frac{f_1^2(2f_1^2+f_2^2/2)}{C_{\!f}^{} \Big(\!\frac{2k_{\!B}^{}T}{\hbar\omega}\!\Big)+\Omega_{N}^{}\Big(\!\frac{\hbar\omega}{2k_{\!B}^{}T}\!\Big)^{\!\lambda}}\!-\!\left(\!\frac{\hbar\omega}{2k_{\!B}^{}T}\!\right)\!\right]\!,
\end{equation}

\begin{multline}\label{eq:C_HO_Qubit_Lambda}
C_{{\rm SS}}^{}(\lambda)\approx\frac{\hbar\omega}{2}\,\frac{f_1^{}f_2^{}(2f_1^2+f_2^2/2)}{C_{\!f}^{} \left(\!\frac{2k_{\!B}^{}T}{\hbar\omega}\!\right)+\Omega_{N}^{}\!\left(\!\frac{\hbar\omega}{2k_{\!B}^{}T}\!\right)^{\!\lambda}}
\\
\times\sqrt{1+\frac{\Omega_{N}^{}}{(2f_1^2+f_2^2/2)^2}\left(\!\frac{\hbar\omega}{2k_{\!B}^{}T}\!\right)^{\!\!1+\lambda}}\,,
\end{multline}
where $\Omega_{N}^{}=\omega^2/N^2$ and $C_{\!f}^{}=(2f_1^2+f_2^2/2)(f_1^2+f_2^2/2)$. The square root in Eq.~\eqref{eq:C_HO_Qubit_Lambda} is the factor $\sqrt{1+\omega^2/\Gamma^2}$ that accounts for $\langle\sigma_y\rangle_{\rm _{SS}}$. The ergotropy follows by substituting Eqs.~\eqref{eq:HO_Qubit_Lambda} and \eqref{eq:C_HO_Qubit_Lambda} into Eq.~\eqref{eq:Ergotropy(lambda)}. 
Although Eq.~\eqref{eq:Approx_linear_distributions} is strictly valid only for $k_{B}^{}T\gg\hbar\omega$, the resulting ergotropy differs from the exact one by less than ten percent already at $k_{\!B}^{}T/\hbar\omega\sim1$.

The bath statistics enters only through the exponent of the term proportional to $\Omega_{N}^{}$, that is, through the finite-size correction $\omega^2/N^2$. For oscillator baths ($\lambda=+1$) this term is suppressed as $\hbar\omega/2k_{\!B}^{}T$ at high temperature, both in the denominator of Eqs.~\eqref{eq:HO_Qubit_Lambda}--\eqref{eq:C_HO_Qubit_Lambda} and under the square root. The ergotropy then approaches the $N$-independent form $A\,\mathcal{C}_0\hbar\omega(\hbar\omega/2k_{\!B}^{}T)$, which is Eq.~\eqref{eq:Ergotropy-Co-&-A=f1-f2} with $\tanh x\to x$. For qubits ($\lambda=-1$) the same term is enhanced as $2k_{\!B}^{}T/\hbar\omega$ and survives at all temperatures when $N$ is finite, reducing the prefactor of the $1/T$ decay by a factor that depends on $\Omega_{N}^{}$. Both reservoirs therefore yield an ergotropy that decays as $1/T$; the qubit reservoir simply does so with a smaller coefficient, as shown in Fig.~\ref{fig:Ergotropy-curves}. As $N\to\infty$ the two coefficients coincide.

\subsection{Bounds on the ergotropy of a qubit}\label{subsec:Bounds}

Since the ergotropy and the coherence of a qubit are both functions of the Bloch components, the positivity of $\rho$, ${\rm{tr}}\{\rho^{2}\}\leq1$, i.e., $|\langle\sigma_x\rangle + i\langle\sigma_y\rangle|^{2}\leq\left(1-\langle\sigma_z\rangle^{2}\right)$, constrains them jointly. In terms of $U$ and the energy-scaled coherence $C$ introduced above, the imposed constraint forces $U^{2}\leq(\hbar\omega/2)^2-\,C^{2}$.
By using such inequality of $U^2$ in $\mathcal{W}=U+\sqrt{U^2+C^2}$ gives an upper bound on ergotropy in terms of the internal energy alone,
\begin{equation}\label{eq:Ergotropy_bounded_by_energy}
    \mathcal{W}(U)\leq\,U^{}+{\hbar\omega}/{2}.
\end{equation}
Similarly, a lower bound can also be obtained by noticing $\mathcal{W}(U)\geq U+|U|$, which yields $\mathcal{W}\geq2\,\text{max}\{0,U\}$~\cite{Choquehuanca-2024}.

On the other hand, eliminating $U$ in favor of the $l_1$-norm $\mathcal{C}$, with $C=(\hbar\omega/2)\mathcal{C}$, gives a bound on ergotropy expressed solely in terms of coherence
\begin{equation}\label{eq:Ergotropy_bounded_by_Coherence}
     \mathcal{W}(\mathcal{C})\leq({\hbar\omega}/2)\big(\sqrt{1-\mathcal{C}^{2}}+1\big).
\end{equation}
The bound Eq.~(\ref{eq:Ergotropy_bounded_by_Coherence}) decreases with coherence because the absolute maximum $\hbar\omega$ is reached by the pure excited state, which has $\mathcal{C}=0$ while a fully coherent qubit state $\mathcal{C}=1$ gives an ergotropy of $\hbar\omega/2$. It does not contradict the fact that, at fixed internal energy, coherence increases the ergotropy. 

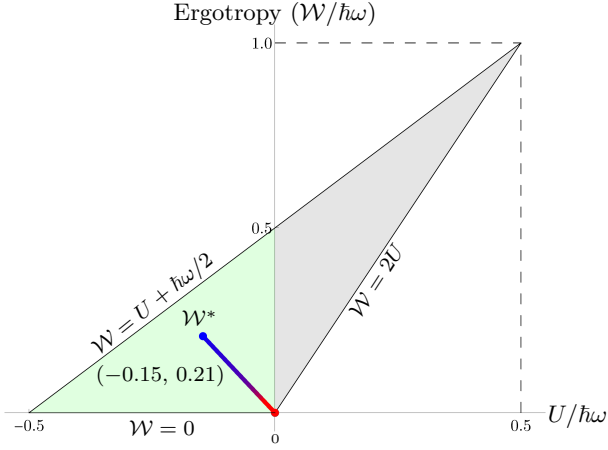
\begin{figure}[t]
    \centering
    \begin{tikzpicture}
        \node (img){\includegraphics[width=0.4\textwidth]{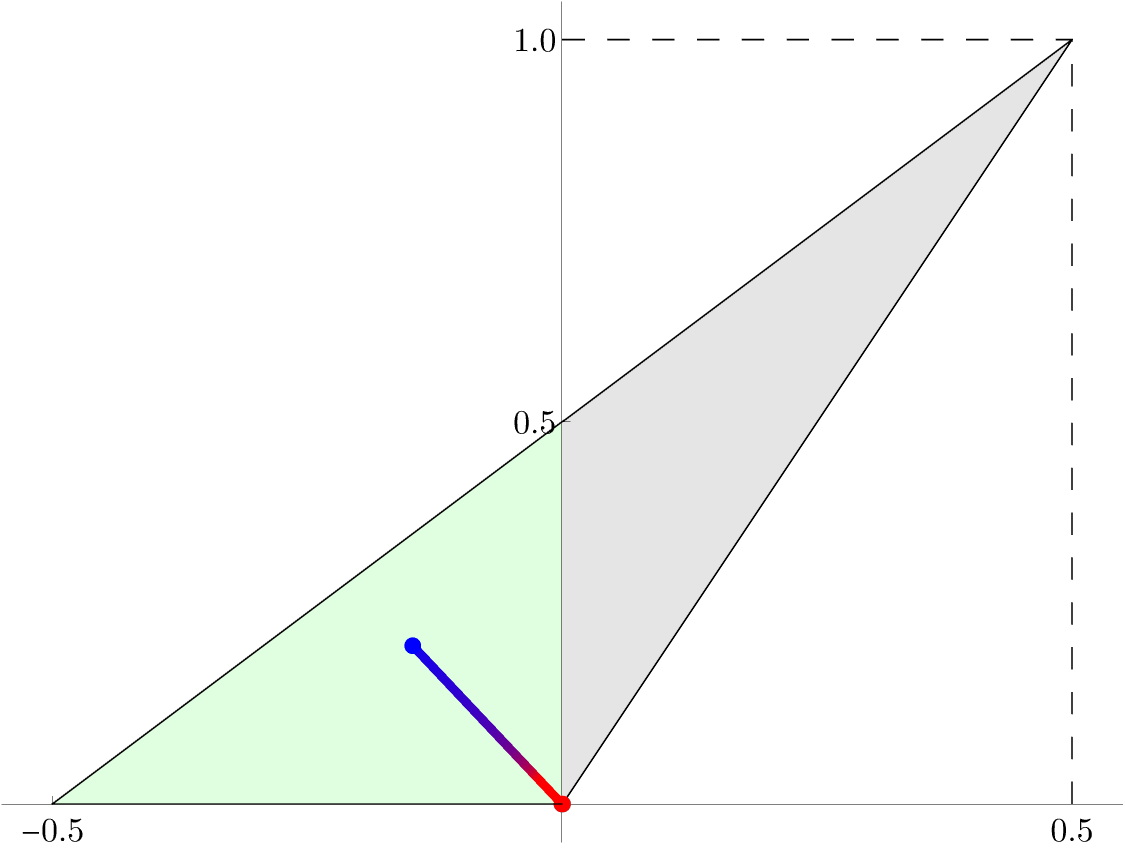}};
        \node[above=of img, node distance=0cm, xshift=0cm, yshift=-1.25cm,]
        {Ergotropy ($\mathcal{W}/\hbar\omega$)};
        \node[above=of img, node distance=0cm, xshift=-1.45cm, yshift=-5cm, rotate=39]
{\footnotesize{$\mathcal{W}=U+\hbar\omega/2$}};
\node[above=of img, node distance=0cm, xshift=1.52cm, yshift=-4.6cm, rotate=57]
{\footnotesize{$\mathcal{W}=2U$}};
\node[above=of img, node distance=0cm, xshift=-1.5cm, yshift=-6.65cm]
{\footnotesize{$\mathcal{W}=0$}};
        \node[above=of img, node distance=0cm, xshift=4cm, yshift=-6.55cm]
{$U/{\hbar\omega}$};
       \node[above=of img, node distance=0cm, xshift=-1cm, yshift=-5.2cm]
{\footnotesize{$\mathcal{W}^*$}};
  \node[above=of img, node distance=0cm, xshift=-1.52cm, yshift=-6cm]
{\footnotesize{$(-{0.15},\,{0.21})$}};
         \node[above=of img, node distance=0cm, xshift=0cm, yshift=-6.75cm]
    {$\scriptscriptstyle{0}$};
    \end{tikzpicture}
    \caption{Allowed ergotropy of a qubit as a function of its internal energy, Eq.~(\ref{eq:Ergotropy_bounded_by_energy}), in units of $\hbar\omega$. The track from blue ($T=0$) to red ($k_{\!B}^{}T\gg\hbar\omega$) is the optimized steady-state $\mathcal{W}_{\rm SS}^*(T)$ of Eq.~(\ref{eq:Optimized_Ergotropy}), with zero-temperature coordinates $(U^{*}/\hbar\omega,\mathcal{W}^{*}/\hbar\omega)=(-0.15,0.21)$.}
    \label{Energy_Bound}
\end{figure}

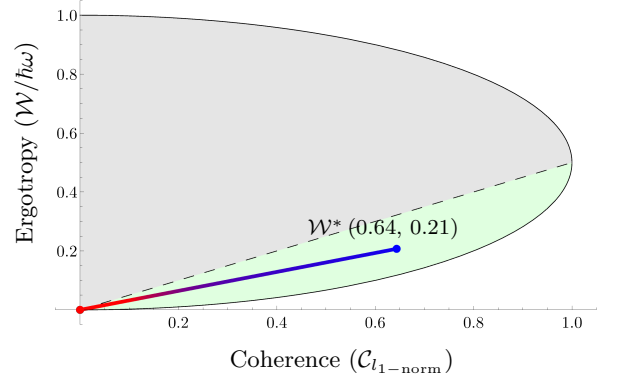
\begin{figure}[t]
    \centering
    \begin{tikzpicture}
        \node (img){\includegraphics[width=0.4\textwidth]{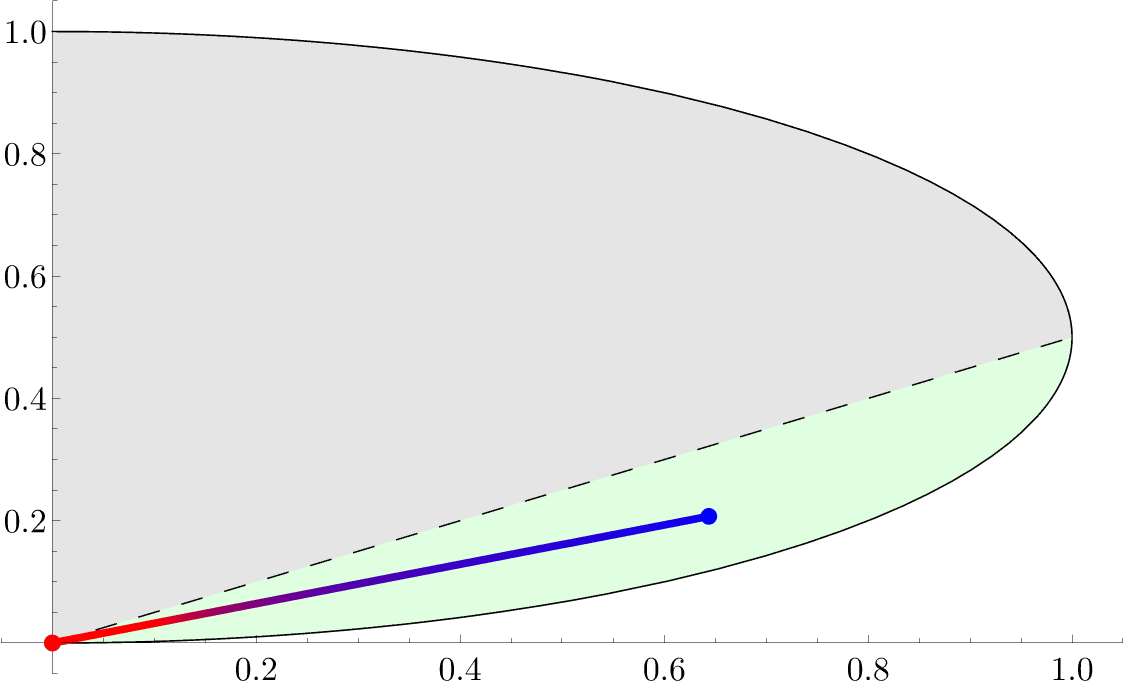}};
        \node[above=of img, node distance=0cm, xshift=-3.7cm, yshift=-3cm, rotate=90,]
        {Ergotropy ($\mathcal{W}/\hbar\omega$)};
        \node[above=of img, node distance=0cm, xshift=0.2cm, yshift=-6.2cm]
        {Coherence ($\mathcal{C}_{l_{\rm1-norm}}$)};
         \node[above=of img, node distance=0cm, xshift=0.75cm, yshift=-4.4cm]
        {\footnotesize{$\mathcal{W}^*\, ({0{.}64},\,{0{.}21})$}};
    \end{tikzpicture}
    \caption{Allowed ergotropy of a qubit as a function of its $l_1$-norm of coherence, bounded by~Eq.~(\ref{eq:Ergotropy_bounded_by_Coherence}) and its lower counterpart. The track from blue ($T=0$) to red ($k_{\!B}^{}T/\hbar\omega\gg1$) is the optimized steady-state ergotropy, $\mathcal{W}_{\rm SS}^*(T)$, with zero-temperature coordinates $(\mathcal{C},\mathcal{W}^{*}/\hbar\omega)=(0.64,0.21)$. Note that the horizontal axis is the dimensionless $l_1$-norm, not the energy-scaled coherence $C=\frac{\hbar\omega}{2}\mathcal{C}\approx0.32\,\hbar\omega$.}
\label{Coherence_Bound}
\end{figure}

The allowed regions are shown as the shaded areas of Figs.~\ref{Energy_Bound} and \ref{Coherence_Bound}, with $\hbar\omega=1$. In Fig.~\ref{Energy_Bound} the internal energy runs from the ground-state value $-1/2$ to the excited-state value $+1/2$. On the right half (gray dark area), $U>0$, the qubit has a population inversion and the ergotropy is at least $2U$. On the left half (green area) there is no population inversion and any nonzero ergotropy is due to coherence. In Fig.~\ref{Coherence_Bound} the region is bounded by an half-ellipse given by $(2\mathcal{W}/\hbar\omega-1)^2+\mathcal{C}^{2}=1$, and the dotted line $\mathcal{W}=(\hbar\omega/2)\mathcal{C}$ separates states with and without population inversion. The origin corresponds to the south pole of the Bloch sphere, the top of the curve to the north pole, and the rightmost point to the equator, where the coherence of a pure state is maximal.

The same constraint can be phrased through the purity $\mathcal{P}\equiv\rm{tr}\{\rho^{2}\}$, $2\mathcal{P}=1+||\langle\vec{\sigma}\rangle||^2$, which turns Eq.~\eqref{eq:Ergotropy_BlochVector} into $\mathcal{W}(\mathcal{P})=({\hbar\omega}/2)\big(\langle\sigma_z\rangle+\sqrt{2\mathcal{P}-1}\big)$. For pure states ($\mathcal{P}=1$) this relation gives $\mathcal{W}=\hbar\omega p_e^{}$ with $p_e^{}=(\langle\sigma_z\rangle+1)/2$, ranging from $0$ at the south pole to $\hbar\omega$ at the north pole. For the maximally mixed state ($\langle\sigma_z\rangle=0$, $\mathcal{P}=1/2$) it vanishes.
Equivalently, since $\det(\rho)=\frac{1}{4}\left(1-|\langle\vec{\sigma}\rangle|^{2}\right)$ and $\mathcal{P}=1-2\det(\rho)$, $\mathcal{W}=({\hbar\omega}/{2})\big(\langle\sigma_z\rangle+\sqrt{1-4\det(\rho)}\big)$.

\section{Thermodynamics of the charged state}\label{sec:thermo}

Every curve in Fig.~\ref{fig:Ergotropy-curves} displays a plateau at low temperature, which implies that the largest ergotropy, attained at $T=0$, remains available over a finite range of bath temperatures. This is attractive for energy storage \cite{Colloquium-QB-2024} and calls for an explanation. In this section we characterize the charged steady state thermodynamically. We first compute its entropy and free energy and compare the latter with the ergotropy (Sec.~\ref{subsec:entropy}); we then show that the plateau is a consequence of the third law of thermodynamics (Sec.~\ref{subsec:nernst}); and finally we compute the work that must be supplied to charge and maintain the state. In the remainder of this section, the subscript ``eq'' refers to the Gibbs (equilibrium) state of the qubit at the bath temperature, which is the steady state for $f_1=0$, and ``neq'' to the coherent (non-equilibrium) steady state $\rho_{_{\rm SS}}^{}$ obtained for $f_1^{}\cdot f_2^{}\neq0$. Unless stated otherwise, the results are given for $N\gg1$ and optimal couplings Eq.~(\ref{eq:OptimizationIdentity}). 

\subsection{Entropy and free energy of the steady state}\label{subsec:entropy}

\begin{figure}[t]
    \centering
  \begin{tikzpicture}
        \node (img){\includegraphics[width=0.45\textwidth]{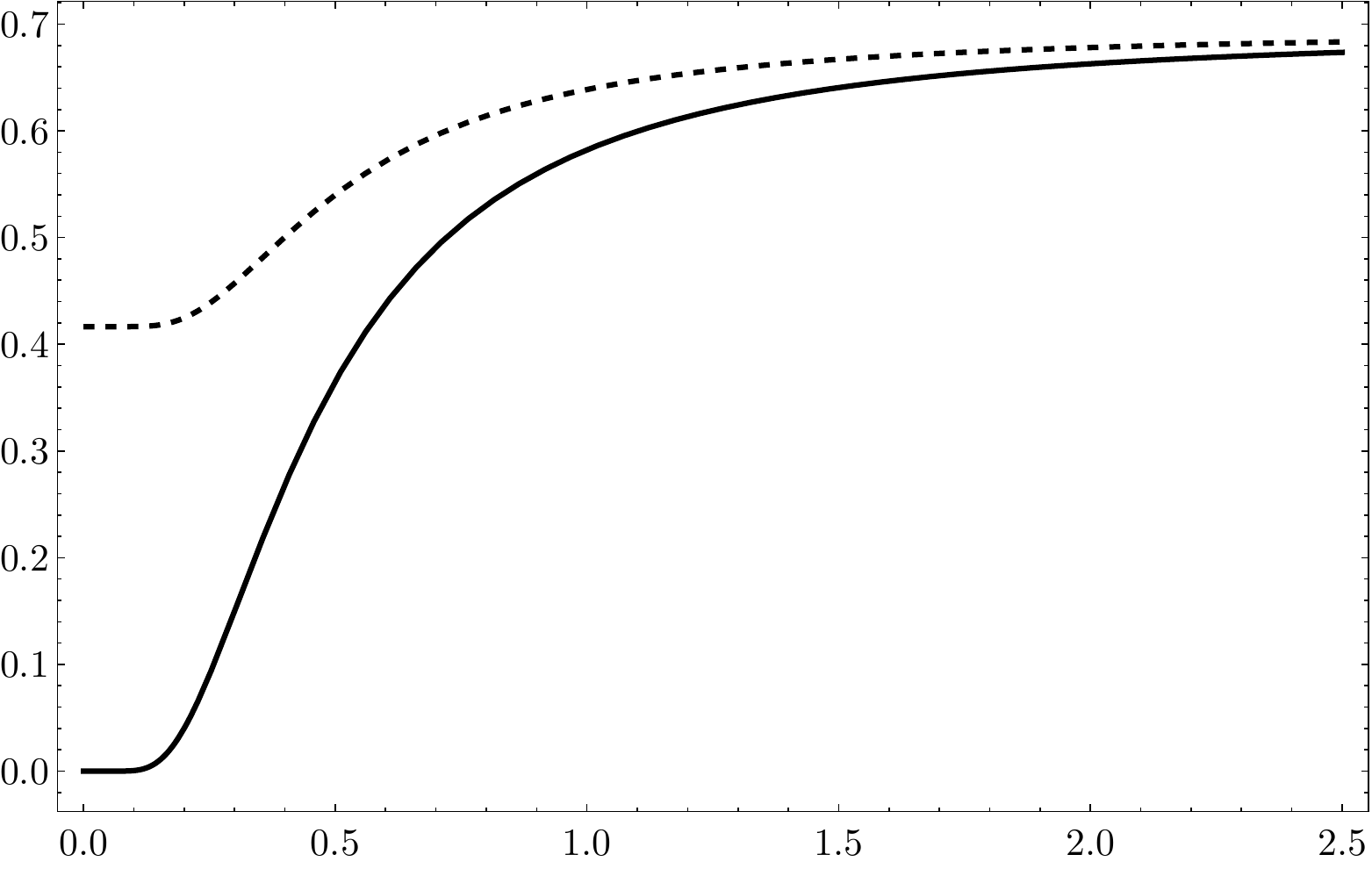}};
        \node[above=of img, node distance=0cm, xshift=-4.2cm, yshift=-3.4cm, rotate=90]
        {Entropy ($S$)};
        \node[above=of img, node distance=0cm, xshift=-1.7cm, yshift=-5cm]
        {\footnotesize{thermal state}};
        \node[above=of img, node distance=0cm, xshift=-2.8cm, yshift=-2.3cm]
        {\footnotesize\shortstack{non-thermal \\steady state}};
        \node[above=of img, node distance=0cm, xshift=0.2cm, yshift=-6.7cm]
        {${k_{\!B}^{}T}/\,{\hbar\omega}$};
        \node[font=\footnotesize] (ln2) at ($(img.north east)+(-0.9cm,-1.2cm)$) {$\ln 2$};
          \draw[->] (ln2.east) -- ($(img.north east)+(-0.3cm,-0.45cm)$);
    \end{tikzpicture}
     \caption{von Neumann entropy for the Gibbs state $S_{\rm eq}$ (solid line), Eq.~(\ref{eq:thermal-S}), and of the coherent (non-thermal) steady state $\rho_{_{\rm SS}}^{}$ (dashed line), Eq.~(\ref{eq:non-eq-S-entropy}), in units of $k_{B}^{}$. While $S_{\rm eq}$ vanishes at $T=0$, $S_{\rm neq}$ tends to $\approx0.416\,k_{B}^{}$ because $\rho_{_{\rm SS}}^{}$ stays mixed. Both have vanishing slope at $T\to0$ and reach the same high-temperature limit.}
    \label{fig:entropies}
\end{figure}
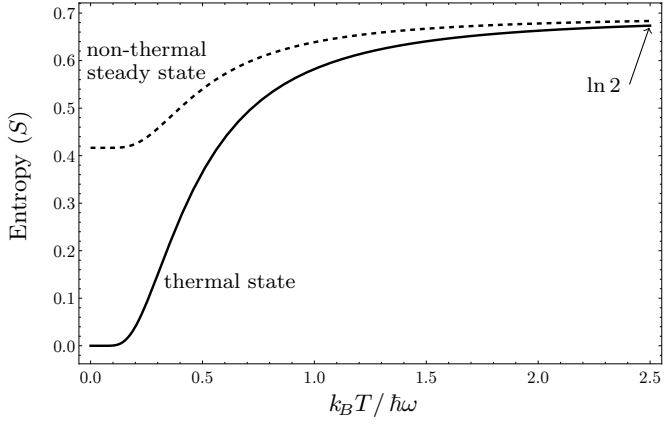

The von Neumann entropy $S(\rho)=-{\rm tr}\{\rho\ln\rho\}$ of a two-level system (qubit) in the Gibbs state $\rho_{\rm eq}^{}$ is \cite{PathriaBeale2011}
\begin{equation}\label{eq:thermal-S}
    S_{\rm eq}^{}=k_{\!B}^{}\{\ln\left[2\cosh(\hbar\omega\beta/2)\right]-(\hbar\omega\beta/2)\tanh(\hbar\omega\beta/2)\}.
\end{equation}
For the non-equilibrium steady state, the eigenvalues of $\rho_{_{\rm SS}}^{}$ are $(1\pm\|\langle\vec{\sigma}\rangle_{_{\rm SS}}\!\|)/2$ with $\|\,\langle\vec{\sigma}\rangle_{_{\rm SS}}\|=\tanh(\hbar\omega\beta/2)/\sqrt{2}$ at the optimum, and the entropy $S(\rho_{_{\rm SS}}^{})$ is
\begin{multline}\label{eq:non-eq-S-entropy}
S_{\rm neq}^{}=-\frac{k_{\!B}^{}}{2}\Biggl\{\ln\left[\frac{1+{\rm sech}^2(\hbar\omega\beta/2)}{8}\right]
\\
+\frac{1}{\sqrt{2}}\tanh(\hbar\omega\beta/2)\ln\left[\frac{1+\frac{1}{\sqrt{2}}\tanh(\hbar\omega\beta/2)}{1-\frac{1}{\sqrt{2}}\tanh(\hbar\omega\beta/2)}\right]\Biggr\}.
\end{multline}
Both entropies are plotted in Fig.~\ref{fig:entropies}. $S_{\rm eq}$ vanishes as $T\to0$, together with its slope, as the thermal state approaches the pure ground state $|g\rangle\langle g|$; this is the Planck form of the third law. $S_{\rm neq}^{}$ also has a vanishing slope at $T\to0$, but it tends to the finite value $S_{\rm neq}(0)=k_{\!B}^{}\,[\tfrac32\ln 2-\tfrac{1}{2\sqrt2}\ln\tfrac{\sqrt2+1}{\sqrt2-1}]\approx0.416\,k_{\!B}^{}$. The non-thermal steady state remains mixed at zero temperature, with purity $3/4$ as noted in Sec.~\ref{subsec:optimum}, because it retains both population and coherence. A residual entropy at $T=0$ is not in conflict with the third law, which is a statement about equilibrium states and the steady state of our model, $\rho_{_{\rm SS}}^{}$, is not one.

\begin{figure}[t]
    \centering
  \begin{tikzpicture}
        \node (img){\includegraphics[width=0.45\textwidth]{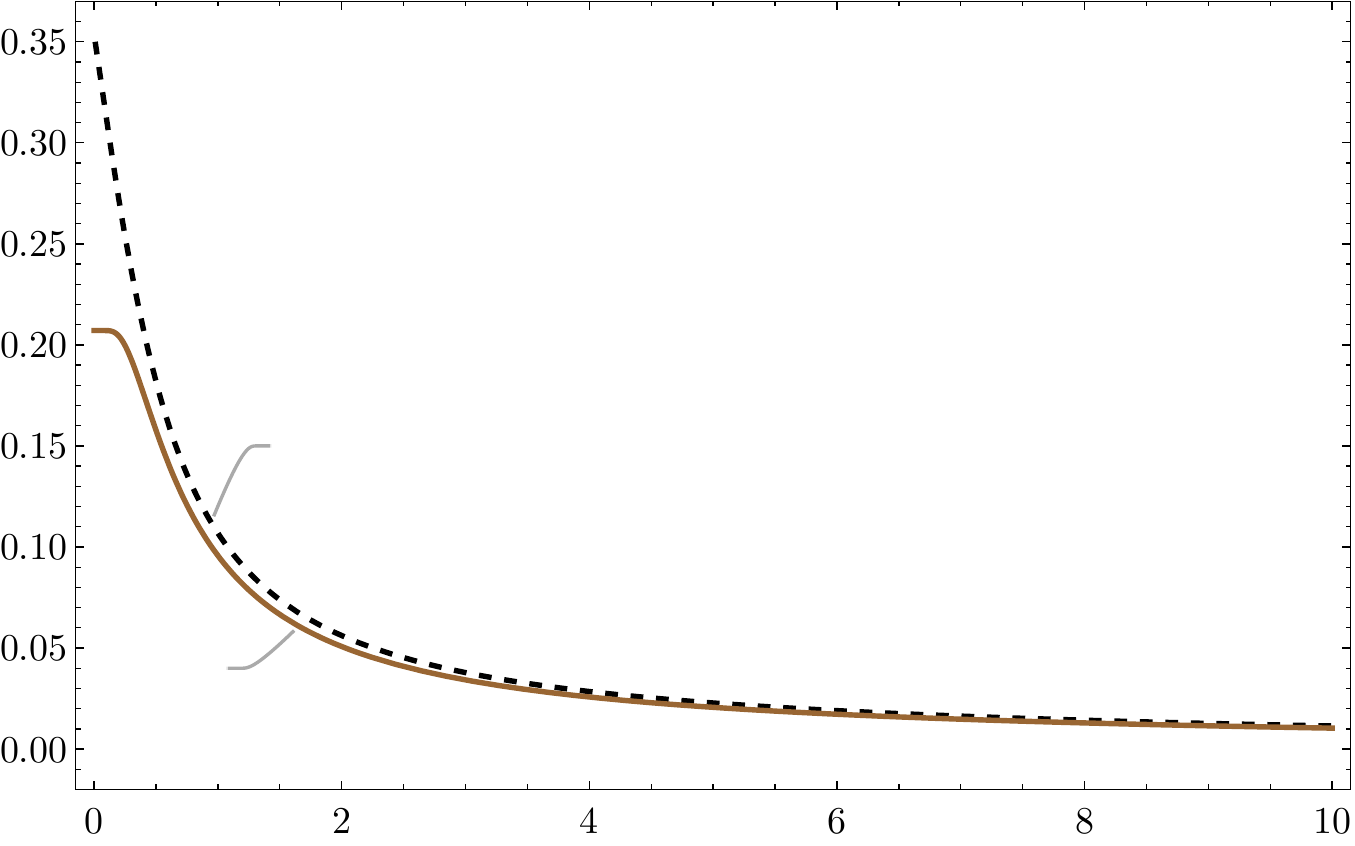}};
        \node[above=of img, node distance=0cm, xshift=-1.1cm, yshift=-4cm]
        {$\Delta F=F_{\rm neq}-F_{\rm eq}$};
        \node[above=of img, node distance=0cm, xshift=-2.9cm, yshift=-5.5cm]
        {$\mathcal{W}_{\rm neq}^{*}$};
        \node[above=of img, node distance=0cm, xshift=0.25cm, yshift=-6.6cm]
        {${k_{\!B}^{}T}/\,{\hbar\omega}$};
    \end{tikzpicture}
     \caption{Nonequilibrium free-energy difference $\Delta F = F_{\rm neq} - F_{\rm eq}$ (black dashed line) and optimized ergotropy $\mathcal{W}_{\rm neq}^{*}$ of Eq.~(\ref{eq:Optimized_Ergotropy}) (brown solid line), in units of $\hbar\omega$. The inequality $\Delta F \ge \mathcal{W}_{\rm neq}$ \cite{Allahverdyan-2004} holds at all temperatures. $\Delta F$ has a clear finite slope at $T=0$, Eq.~(\ref{eq:slope-DeltaF}), whereas $\mathcal{W}_{\rm neq}^*$ is flat.} 
    \label{fig:DeltaF-ergotropy}
\end{figure}
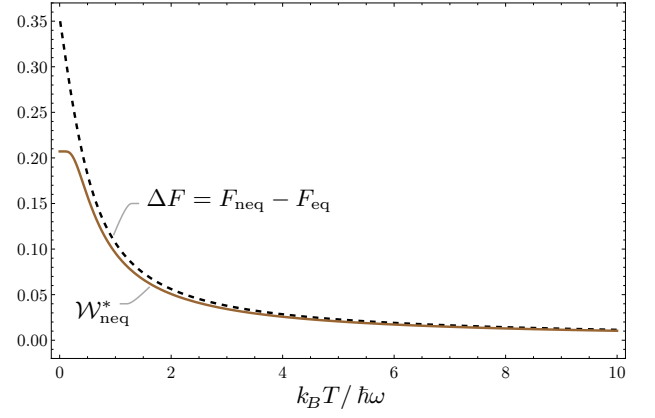

The non-equilibrium free energy, $F(\rho)\!=\!\langle\mathcal{H}(\rho)\rangle\!-\!TS(\rho)$\cite{GLandi-entropy-production}, with $T$ the bath temperature, quantifies the distance of $\rho_{\rm neq}^{}$ from the thermal state $\rho_{\rm eq}^{}$~\cite{Funo_2018}. The equilibrium free energy, $F(\rho_{\rm eq}^{})=U_{\rm eq}-T S_{\rm eq}$,  is~\cite{PathriaBeale2011}
\begin{equation}\label{eq:thermal-F}
F_{\rm eq}=-k_{\!B}^{}T\ln\left[2\cosh(\hbar\omega\beta/2)\right].
\end{equation}
For the non-thermal steady state, $\rho_{_{\rm SS}}^{}$, we obtain 
\begin{equation}\label{neq_free_energy}
F_{\rm neq}^{}\equiv F(\rho_{_{\rm SS}}^{})= U_{\rm neq}^{}-T S_{\rm neq}^{}\,,
\end{equation}
where $U_{\rm neq}\!=\!(\hbar\omega/2)(\mathcal{C}_0{f_1^{}}{f_2^{-1}}-1)\tanh({\hbar\omega}/{2k_{B}^{}T})$, with the optimized value $U_{\rm neq}=-\mathcal{W}_{\rm SS}^*/\sqrt{2}$. It is clear that $U_{\rm neq}\rightarrow U_{\rm eq}$ when $\mathcal{C}_0^{}\rightarrow 0$. The difference $\Delta F\equiv F_{\rm neq}-F_{\rm eq}$ is the maximum work extractable from $\rho_{_{\rm SS}}^{}$ when the bath itself may be used as a resource; the ergotropy is the maximum extractable by unitary operations on the qubit alone, and therefore $\Delta F\geq\mathcal{W}_{\rm neq}\geq0$ \cite{Allahverdyan-2004}. Figure~\ref{fig:DeltaF-ergotropy} confirms this inequality at all temperatures. The two quantities coincide in the high-temperature regime, where both decay as $1/T$, and differ most at low temperature, where $\Delta F(0)=U_{\rm neq}(0)-U_{\rm eq}(0)=\hbar\omega/(2\sqrt2)\approx0.35\,\hbar\omega$ while $\mathcal{W}^{*}(0)\approx0.21\,\hbar\omega$.

Figure~\ref{fig:DeltaF-ergotropy} shows a remarkable clear difference between the two curves at low temperature. $\Delta F$ has a finite slope at $T=0$, while $\mathcal{W}_{\rm neq}$ is flat. Since $\partial_T^{}U_{\!\rm eq}$, $\partial_T^{}U_{\rm neq}$ and $S_{\rm eq}$ all vanish as $T\to0$, differentiating $F_{\rm neq}^{}-F_{\rm eq}^{}$ gives
\begin{equation}\label{eq:slope-DeltaF}
     \lim_{T\to0}\Big(\frac{\partial\Delta F}{\partial T}\Big)=-S_{\rm neq}(0)\approx-0.42\,k_{\!B}^{}.
\end{equation}
The finite slope of $\Delta F$ is thus a direct consequence of the residual entropy of the non-thermal state with steady-state coherence $\rho_{_{\rm SS}}^{}$.

\subsection{Low-temperature plateau and the third law}\label{subsec:nernst}

The third law of thermodynamics, in the form of Nernst's postulate, states that the entropy change of an isothermal process between equilibrium states vanishes as the temperature approaches to zero, $\lim_{T\to0}^{}\Delta S=0~\cite{StrasbergBook}$. For a process at constant pressure, $\Delta G=\Delta {\rm H}-T\Delta S$, and the postulate can be written as~\cite{Callen1985}
\begin{equation}\label{eq:Limit of G and H}
\lim_{T \to 0}\frac{\Delta {\rm H}-\Delta G}{T}= \lim_{T \to 0} \Delta S = 0,
\end{equation}
so that $\Delta {\rm H}-\Delta G$ must vanish faster than $T$. A well-known consequence is that the heat capacity $C_P^{}=(\partial {\rm H}/\partial T)_P^{}=T(\partial S/\partial T)_P^{}$ vanishes at absolute zero, and with it the low-temperature slope of the equilibrium enthalpy. This is the property we need.

For the qubit considered here there is no mechanical volume degree of freedom and, consequently, no $PV$ work contribution. Thus, the enthalpy ${\rm H}=U+PV$ and Gibbs potential $G=U-TS+PV$ \cite{Callen1985} reduce to the internal energy $U={\rm tr}\{\rho H\}$ and the nonequilibrium free energy $F(\rho)$ of Sec.~\ref{subsec:entropy}, respectively. Nevertheless, we retain the thermodynamic notation ${\rm H}$ and $G$ in what follows to make direct contact with the standard formulation of the Nernst postulate in Eq.~\eqref{eq:Limit of G and H}. For the Gibbs state,
\begin{equation}\label{eq:thermal-H}
    {\rm H}_{\rm eq} = -({\hbar\omega}/{2})\tanh(\hbar\omega\beta/2),
\end{equation}
whose derivative $\partial_T {\rm H}_{\rm eq}=k_{\!B}^{}(\hbar\omega\beta/2)^2\,{\rm sech}^2(\hbar\omega\beta/2)$ ~\cite{PathriaBeale2011} is the heat capacity of the qubit and vanishes exponentially as $T\to0$. For the non-equilibrium state, $\langle\sigma_z\rangle_{_{\rm SS}}^{^{\!N\gg1}}$ gives
\begin{equation}\label{eq:non-eq-enthalpy-Hth}
    {\rm H}_{\rm neq} = \left(1 - \mathcal{C}_{0}^{}{f_1^{}}{f_2^{-1}}\right) {\rm H}_{\rm eq}.
\end{equation}
The key observation is that the ergotropy of Eq.~\eqref{eq:Ergotropy-Co-&-A=f1-f2} is itself proportional to the \emph{equilibrium} enthalpy, $\mathcal{W}_{\rm neq} = -2A \mathcal{C}_0 {\rm H}_{\rm eq}$, with a proportionality constant that does not depend on temperature. For the optimized case [see Eq.~\eqref{eq:Optimized_Ergotropy}], $\mathcal{W}_{\rm neq} = (1 - \sqrt{2}) {\rm H}_{\rm eq}$.  Hence
\begin{equation}\label{eq:Nernst-limits}
\lim_{T\to 0}\frac{\partial G_{\rm eq}}{\partial T}=\lim_{T\to 0}\frac{\partial {\rm H}_{\rm eq}}{\partial T}=0
\Longrightarrow
\lim_{T\to 0}\frac{\partial\mathcal{W}_{\rm neq}}{\partial T}=0 .
\end{equation}
The plateau of the ergotropy is therefore inherited from the Nernst's heat theorem. When the system is in contact with the bath elements, its extractable work is a fixed multiple of the bath's equilibrium enthalpy, and that enthalpy is flat at low temperature because the heat capacity vanishes (Schottky anomaly). The argument has a general part, the vanishing of $C_P^{}$, and a model-specific part, the temperature-independent proportionality between $\mathcal{W}_{\rm neq}$ and ${\rm H}_{\rm eq}$; the latter holds in the large-cluster limit. For finite $N$ the temperature enters Eq.~\eqref{eq:Ergotropy(lambda)} only through $n(\lambda)$ and $\tanh(\hbar\omega\beta/2)$, both of which have exponentially small derivatives at $T\to0$, so the plateau persists for all $N$ even though the simple proportionality does not.

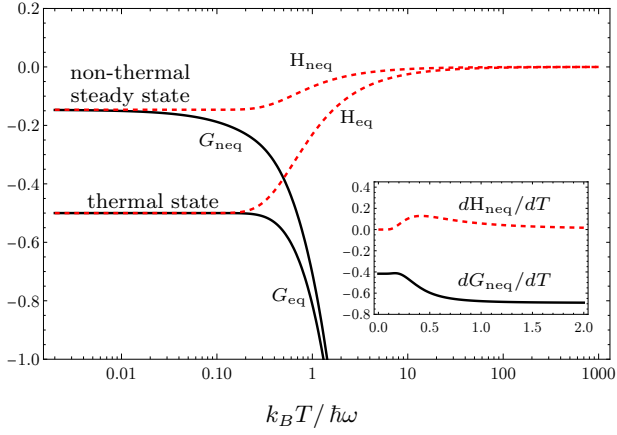
\begin{figure}[t]
    \centering
  \begin{tikzpicture}
        \node (img){\includegraphics[width=0.45\textwidth]{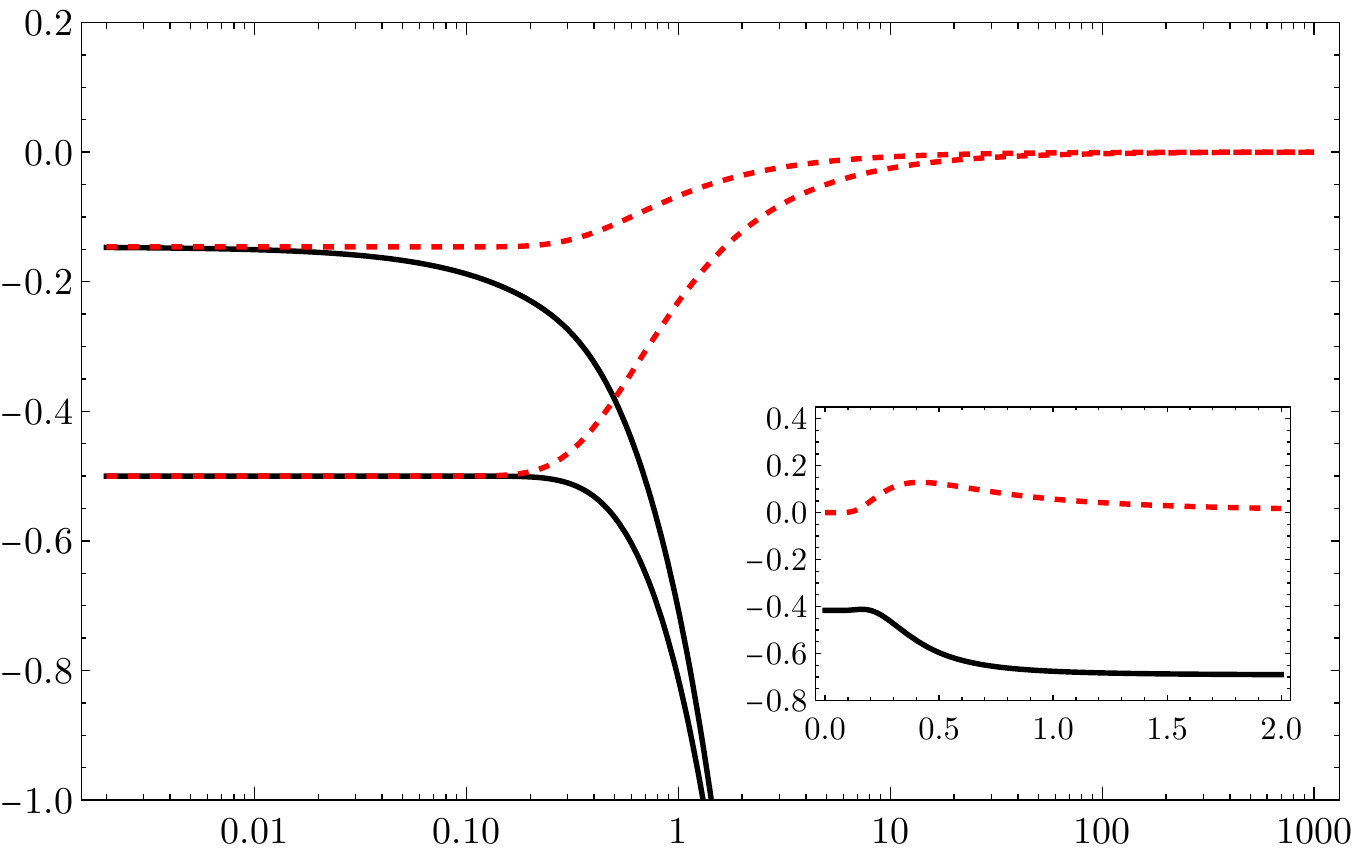}};
        \node[above=of img, node distance=0cm, xshift=-2.1cm, yshift=-4cm]
        {\footnotesize{thermal state}};
         \node[above=of img, node distance=0cm, xshift=-2.4cm, yshift=-2.65cm]
        {\footnotesize\shortstack{non-thermal \\steady state}};
          \node[above=of img, node distance=0cm, xshift=0.6cm, yshift=-2.95cm]
        {\scriptsize{${\rm H}_{\rm eq}$}};
         \node[above=of img, node distance=0cm, xshift=0cm, yshift=-2.2cm]
        {\scriptsize{${\rm H}_{\rm neq}$}};
        \node[above=of img, node distance=0cm, xshift=-0.3cm, yshift=-5.3cm]
        {\scriptsize{$G_{\rm eq}$}};
         \node[above=of img, node distance=0cm, xshift=-1.2cm, yshift=-3.25cm]
        {\scriptsize{$G_{\rm neq}$}};
         \node[above=of img, node distance=0cm, xshift=2.55cm, yshift=-4.1cm]
        {\scriptsize{$d{\rm H}_{\rm neq}/dT$}};
        \node[above=of img, node distance=0cm, xshift=2.55cm, yshift=-5.1cm]
        {\scriptsize{$dG_{\rm neq}/dT$}};
        \node[above=of img, node distance=0cm, xshift=0cm, yshift=-6.9cm]
        {${k_{B}^{}T}/\,{\hbar\omega}$};
    \end{tikzpicture}
 \caption{Enthalpies (red dashed lines) and Gibbs potentials (black solid lines) of the Gibbs state (${\rm H}_{\rm eq}, G_{\rm eq}$) and of the coherent (non-thermal) steady state (${\rm H}_{\rm neq}, G_{\rm neq}$), in units of $\hbar\omega$. ${\rm H}_{\rm eq}$, $G_{\rm eq}$ and ${\rm H}_{\rm neq}$ have vanishing slope as $T\to0$; $G_{\rm neq}$ does not, with $\lim_{T\to 0}^{}\partial_T{}G_{\rm neq}\approx -0.416\,k_{\!B}^{}$ (inset).}
    \label{fig:Gibbs-vs-enthalpies}
\end{figure}

Figure~\ref{fig:Gibbs-vs-enthalpies} compares the four thermodynamic potentials. While ${\rm H}_{\rm eq}$, $G_{\rm eq}$ and ${\rm H}_{\rm neq}$ are flat at low temperature, $G_{\rm neq}=F_{\rm neq}$ is not and its slope at $T=0$ is $-S_{\rm neq}(0)\approx-0.416\,k_{\!B}^{}$ (inset) which is the same residual entropy responsible for Eq.~\eqref{eq:slope-DeltaF}. The ergotropy sides with the enthalpies, not with the free energies, because it does not depend on the entropy of the state.

Finally, we address what sustains the coherent steady state. Although the master equation of Eq.~\eqref{eq:GenericMasterEquation} is time independent, it descends from a collision model in which the interaction is switched on and off at every collision, and it has been shown that such boundary-driven master equations are thermodynamically consistent only once the work performed by the switching is accounted for~\cite{Barra2015,Q-info-Unifying-Frame-Massimi-Esposito-2017,G.Landi-De-Chiara-Weakly-Collisional-model-2019,Landi_RMP_2022}. For the composite interaction of Eq.~\eqref{eq:Vi-General-Interaction-Hamiltonian} at resonance, the work rate for arbitrary cluster size and bath statistics is (Appendix~\ref{Apx:Cost})
\begin{equation}\label{eq:Wdot-general}
  \dot{W}=f_1^{}\hbar\omega\Big(
    f_1\big\langle[B_{\lambda}^{},B_{\lambda}^{\dagger}]\big\rangle
    -\frac{f_2^{}}{2}\big\langle\{B_{\lambda}^{},B_{\lambda}^{\dagger}\}\big\rangle
     \langle\sigma_x\rangle\Big),
\end{equation}
which for $N=1$ reproduces the results of \cite{R-Ancheyta-Enhanced-2021} for qubit and oscillator baths. Two properties follow at once. The rate is proportional to $f_1^{}$ and vanishes identically for $f_1^{}=0$, which is exactly the case in which the steady state is a Gibbs state with zero ergotropy. The parallel coupling ($f_1^{}$) is the port through which work enters the battery, and the orthogonal coupling ($f_2^{}$) the channel that converts it into extractable work. At $T=0$ the ancillas arrive in their ground state and can only absorb energy, so the charging shown in Fig.~\ref{fig:Ergotropy_1Q-system} is paid for entirely by switching the interaction with the environmental ancillas. In the steady state of the system, first law of thermodynamics dictates that $dU/dt=\dot{Q}+\dot{W}=0$~\cite{Kosloff_2013}, i.e., $\dot{W}=-\dot{Q}$, and therefore, this work is dissipated into the stream of ancillas as heat at the same rate.

In the large-cluster limit, Eq.~\eqref{eq:Wdot-general} reduces to
\begin{equation}\label{eq:Wdot-largeN}
  \dot{W}^{^{N\gg 1}}_{{\rm SS}}
  \!=\!\hbar\omega N\big[1\!+\!(\lambda\!+\!1)n(\lambda)\big]
   \frac{f_1^{4}}{f_1^{2}+f_2^{2}/2}
   \tanh\!\left(\!\frac{\hbar\omega}{2k_{\!B}^{}T}\!\right).
\end{equation}
Unlike the ergotropy \eqref{eq:Ergotropy-Co-&-A=f1-f2}, which becomes independent of the bath statistics in this limit, the cost of maintaining the charged state does not. The $\lambda$-dependent prefactor of the above expression is $N$ for qubits and $N\coth(\hbar\omega\beta/2)$ for oscillators. The two reservoirs deliver the same stored work at different prices, the oscillator bath becoming more expensive relative to the qubit bath setting, as the temperature grows.

\section{Scaling the qubit battery}\label{subsec:Battery-Scaling}

We now let the battery consist of $k$ noninteracting qubits, $k=1,\dots,7$, with Hamiltonian $\tilde{\mathcal{H}}_{S}^{}=(\hbar\omega/2)\sum_{i=1}^{k}\sigma_z^{(i)}$, all of which interact with the same incoming cluster through Eq.~\eqref{eq:Vi-General-Interaction-Hamiltonian}. The system operator in the master equation \eqref{eq:GenericMasterEquation} becomes $\tilde{s}=f_1^{}\sum_i\sigma_z^{(i)}+f_2^{}\sum_i\sigma_-^{(i)}$, and the unitary part $-i[\tilde{\mathcal{H}}_{S}^{},\rho]$; the bath averages are unchanged. The steady state is obtained numerically by vectorizing the $2^k\times2^k$ density matrix into a $2^{2k}$-component vector and finding the kernel of the corresponding $2^{2k}\times2^{2k}$ Liouvillian, at $T=0$, $N\gg1$, and the couplings of Eq.~\eqref{eq:OptimizationIdentity}.

\begin{figure}[t]
\includegraphics[width=0.4\textwidth]{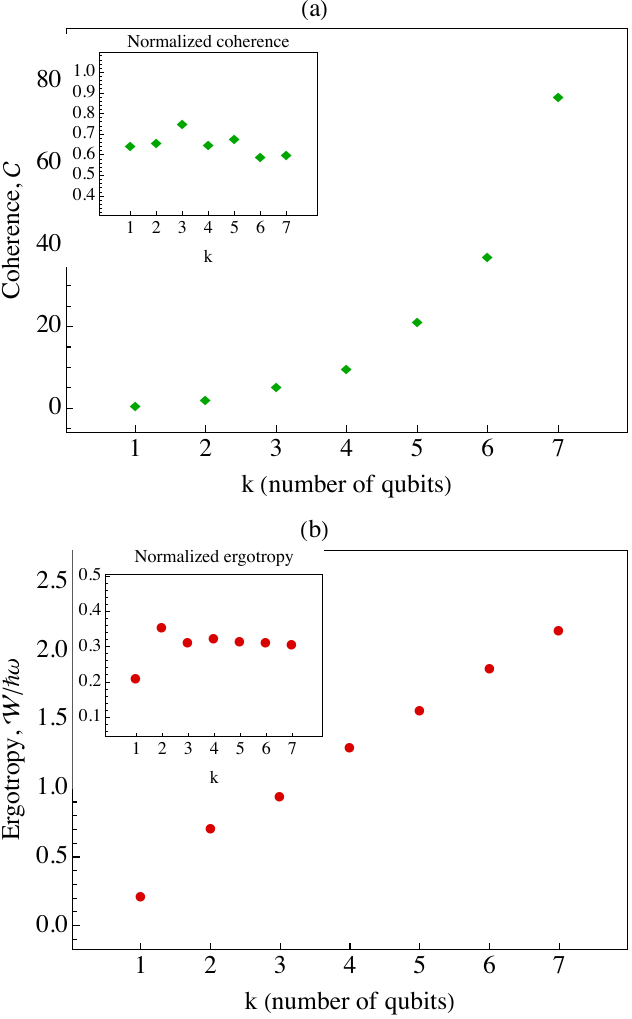}
\caption{(a) Scaling behavior of the $l_1$-norm of coherence and its normalized version (inset), and (b) stored ergotropy as a function of the number of qubits in the system. While the $\mathcal{C}$ shows an exponential scaling as the number of qubits are increased in the system, the ergotropy grows linearly. Note that, the properly normalized $\mathcal{C}$ according to the system size follows a rather horizontal trend with the number of qubits.}
\label{fig:Scaling}
\end{figure}

Figure~\ref{fig:Scaling} shows the $l_1$-norm of coherence and the ergotropy of the steady state as functions of $k$. We observe that the two quantities scale very differently. The $l_1$-norm of coherence grows quite fast with the Hilbert space, well described by the fit $0.6(2^k-1)$, while the ergotropy grows linearly, adding $\approx0.3\,\hbar\omega$ per qubit. The difference between the two is one of normalization. The $l_1$-norm adds up the off-diagonal elements of the density matrix, of which a $k$-qubit state has $2^k(2^k-1)$, and its maximum value is $2^k-1$, which grows with the dimension of the Hilbert space~\cite{Coherence-l1norm-2014}. On the other hand, the ergotropy is bounded by the energy gap $k\hbar\omega$ per qubit, which grows with the number of qubits in the system. However, when normalized with respect to their respective maxima both quantities are nearly flat. $\mathcal{C}/(2^k-1)$ stays between $0.6$ and $0.75$ [inset of Fig.~\ref{fig:Scaling}(a)], and $\mathcal{W}/k\hbar\omega\approx0.31$ for $k\geq2$ [inset of Fig.~\ref{fig:Scaling}(b)]. The battery therefore stores a fixed fraction of its capacity at every size, and the exponential growth of the coherence reflects the growth of the space it lives in rather than an increase in extractable work. The same hierarchy, $\mathcal{C}_{l_1}\sim2^k$ against $\mathcal{W}\sim k$, was found for qubits coupled collectively to a thermal field \cite{Coherence-Cakmak-2020}, since most of the exponentially many coherences sit in degenerate energy subspaces where no unitary can convert them into work.

\section{Conclusions}\label{sec:conclusions}

We have studied a qubit as a minimal model of a quantum battery charged by a thermal reservoir. The qubit interacts repeatedly with clusters of $N$ thermal ancillas, either qubits or oscillators, through the composite interaction \eqref{eq:Vi-General-Interaction-Hamiltonian}, whose parallel and orthogonal components together generate coherence in the energy basis of the steady state. These steady-state coherences are responsible for the finite ergotropy of the battery such that at the steady state the qubit has no population inversion, so all of its ergotropy is due to coherences. We determined the ratio of composite couplings that maximizes the stored work, $2f_1^{2}=(1+\sqrt{2})f_2^{2}$, for which the ergotropy in the large-cluster limit is $2\mathcal{W}^{*}=(\sqrt{2}-1){\hbar\omega}\tanh(\hbar\omega/2k_{\rm B}^{}T)$. Its zero-temperature value coincides with that of a qubit battery charged by a resonant coherent drive \cite{Morrone-2023}. However, in the present model the same amount of ergotropy is obtained from an engineered interaction with a thermal bath, a low-cost thermodynamic resource, and we are able to analyze its temperature dependence in full.

Our analytical solutions depend on the bath only through a single parameter $\lambda$ that encodes its quantum statistics, so that reservoirs of qubits and of oscillators are treated at once. For finite clusters the two types of ancillas differ with the ergotropy decaying more slowly with temperature for oscillators. On the other hand, in the large-cluster limit the ergotropy loses its dependence on $\lambda$ altogether, while the work required to sustain the steady-state does not. This implies that the two reservoirs store the same ergotropy at different work costs.

We also extended the battery to $k$ noninteracting qubits that interact with the same incoming clusters. The $l_1$-norm of coherence of the steady state grows exponentially with $k$, while the ergotropy grows linearly, adding about $0.3\,\hbar\omega$ per qubit. The contrast reflects the quantities' different maxima rather than a change in how the battery charges, since the largest coherence a $k$-qubit state can hold, $2^k-1$, grows with the dimension of the Hilbert space, whereas the largest ergotropy, $k\hbar\omega$, grows with the number of levels. Normalized to these maxima, both quantities are nearly independent of $k$, at roughly $0.6$ and $0.3$ respectively. The battery thus stores a fixed fraction of its capacity at every size, and the exponentially many coherences of the larger systems, most of which sit in degenerate energy subspaces, do not translate into additional extractable work.

We also derived the bounds that internal energy and coherence impose on the ergotropy of a qubit, both saturated by pure states. The largest ergotropy, $\hbar\omega$, belongs to the fully inverted state and carries no coherence, while the most coherent pure state stores half of it. The optimized steady state, with $\mathcal{W}_{\rm SS}^{*}=(\sqrt{2}-1)\hbar\omega/2$ at $T=0$, reaches about a fifth of the maximum and lies on the noninverted side of the allowed region, which is a geometric statement that its ergotropy is of coherent origin.

We identified a way to understand the plateau displayed by all ergotropy curves at low temperature in the context of the third law of thermodynamics, as stated by the Nernst postulate. The optimized ergotropy is a fixed multiple of the equilibrium enthalpy of the qubit, whose slope is the heat capacity and vanishes as $T\to0$. For $k_{\rm B}^{}T\ll\hbar\omega$ the thermal population of the excited state is exponentially small, the equilibrium energy is insensitive to temperature, and the stored ergotropy inherits this insensitivity. The non-equilibrium steady state that holds this ergotropy nevertheless retains a finite entropy at zero temperature, so that its free energy is not flat. The plateau is a property of the enthalpy, not of the free energy, and it freezes the stored ergotropy over a considerable range of temperatures.

\section{Acknowledgments}
R.R.-A. thanks DGAPA-UNAM, Mexico for support under Project No. IA101826. O.S.Y.-S. thanks SECIHTI, Mexico for his PhD Scholarship. B.\c{C}. is partially supported by the Farmingdale State College Provost's Office Summer Research Award Program.


\appendix

\section{Gibbs states for the bath element}\label{Apx:Reservoir}

For a bath of two-level sytems (qubits) with Hamiltonian $\mathcal{H}_{B}^{}=\hbar\omega\sigma_z^{_B}/2$, the thermal state is
$\rho_{q}^{_B}=\exp(-\hbar\omega\beta\sigma_z^{_B}/2)/{Z}$, where $Z=2\cosh({\hbar\omega\beta}/{2})$.
For oscillator bath elements we have $\mathcal{H}_{\!B}^{}=({\hbar\omega}/{2})(b^{^\dagger}b+b\,b^{^\dagger})$, where $b$ ($b^{^\dagger}$) is the standard annihilation (creation) operator, the Gibbs state is $\rho_{\rm _O}^{_B}=\sum_nP_n|n\rangle\langle n|$,
with Boltzmann weights on the Fock states $n=0,1,\dots$, $P_n={e^{-\beta E_n}}/\left({\sum_{m}e^{-\beta E_m}}\right)$, and $E_m=\hbar\omega(m+1/2)$.

\subsection{Expectation values in the Gibbs states}\label{Apx-sub:ValuesForThe:thermal-states}

Depending on $\lambda$, the master equation (\ref{eq:GenericMasterEquation}) involves $\langle \sigma_-^{B}\sigma_+^{B}\rangle$ and $\langle\sigma_+^{B}\sigma_-^{B}\rangle$, or $\langle bb^{^\dagger}\rangle$ and $\langle b^{^\dagger}b\rangle$ evaluated in the thermal states $\rho_{q}^{_B}$ and $\rho_{\rm _O}^{_B}$, respectively. The corresponding commutator and anticommutator are listed in Table~\ref{table:Expected_values_Qbits-QHO}.

\begin{table}[t]
    \centering
 {  \renewcommand{\arraystretch}{1.8}
    \begin{tabular}{|c|c|}\hline
         Two-level system & Harmonic oscillator\\\hline
         $\begin{array}{rcl}
    \langle\{\sigma_-^{_B},\sigma_+^{_B}\}\rangle&=&1\\
    \langle\left[\sigma_-^{_B},\sigma_+^{_B}\right]\rangle&=&\tanh\left(\hbar\omega\beta/2\right)
         \end{array}$
         &
        $\begin{array}{rcl}
    \langle\{b,b^{^\dagger}\}\rangle&=&\coth\left(\hbar\omega\beta/2\right)\\
         \langle[b,b^{^\dagger}]\rangle&=&1
         \end{array}$\\
         \hline
    \end{tabular}
    }
     \caption{Thermal expectation values of the commutator and anticommutator of the ladder operators of a single ancilla, for qubits and harmonic oscillators.}
    \label{table:Expected_values_Qbits-QHO}
\end{table}

Following \cite{Sargsyan-2021}, who introduced a parameter $\varepsilon$ in the commutation relation $aa^{{\dagger}} -\varepsilon \, a^{{\dagger}}a=1$ to treat fermions ($\varepsilon=-1$) and bosons ($\varepsilon=1$) at once, the entries of Table~\ref{table:Expected_values_Qbits-QHO} can be written in terms of the thermal occupations,
\begin{equation}
    \begin{split}
            \tanh\left(\hbar\omega\beta/2\right)&=1-2n_{{\rm F}}^{},
            \\
             \coth\left(\hbar\omega\beta/2\right)&=1+2n_{{\rm B}}^{},
    \end{split}
\end{equation}
with $n_{{\rm F}}^{}$ and $n_{{\rm B}}^{}$ the Fermi-Dirac and Bose-Einstein distributions.

\subsection{Generic bath elements}\label{Apx-sub:Generic-Baths}

We adopt the same idea, with the parameter $\lambda$ also used in \cite{Palafox-Heat-2022}, and define generic ancilla ladder operators $b_{\lambda}^{}$, $b_{\lambda}^{\dagger}$ through
\begin{equation}\label{eq:ConmmutationRelation}
    b_{{\lambda}}^{} b^{{\dagger}}_{{\lambda}} -\lambda \,
b^{{\dagger}}_{{\lambda}} b_{{\lambda}}^{}=1 ,
\end{equation}
which is the bosonic commutator for $\lambda=1$ and the qubit anticommutator for $\lambda=-1$. 
In the Gibbs state,
\begin{equation}\label{eq:generic_bb_lambda}
    \begin{split}
    \langle b^{{\dagger}}_{{\lambda }}b_{{\lambda}}^{}\rangle &= n(\lambda),
    \\
    \langle b_{\lambda}^{} b^{\dagger}_{\lambda}\rangle &=1+\lambda\,n(\lambda),
    \end{split}
\end{equation}
with $n(\lambda)=[\exp{\left(\hbar\omega/k_{\rm B}^{}\, T\right)}-\lambda]^{-1}$, 
and hence
\begin{equation}
    \begin{split}
      \langle[ b_{\lambda}^{}, b^{\dagger}_{\lambda} ] \rangle&=1+(\lambda-1)\,n(\lambda),
      \\
      \langle\{b_{\lambda}^{}, b^{\dagger}_{\lambda} \}\rangle&=1+(\lambda+1)\,n(\lambda).   
    \end{split}
\end{equation}
For a cluster of $N$ noninteracting ancillas we define the collective operator
\begin{equation}\label{eq:GenericBath}
B_{\lambda}^{} ={\sum}_{i=1}^{N}b^{{(i)}}_{\lambda}\, ,
\end{equation}
which for qubits is the collective spin operator
\begin{equation}\label{eq:generic_bath_S}
    S_{\pm}^{}\equiv{\sum}_{j=1}^N\sigma_{\pm}^{{(j)}}\ ,
\end{equation}
and for oscillators
\begin{equation}\label{eq:generic_bath_B}
            B\equiv{\sum}_{k=1}^Nb_{k}^{} \qquad  {\rm and} \qquad B_{}^\dagger\equiv{\sum}_{k=1}^Nb_{k}^{\dagger}\, .
\end{equation}

\begin{table}[t]
    \centering
{  \renewcommand{\arraystretch}{1.8}
   \begin{tabular}{|c|c|}\hline
         $N$ QUBITS & $N$ OSCILLATORS\\\hline
         $\lambda=-1$ & $\lambda=1$\\
         $\begin{array}{rcl}
\langle \{B_{\lambda}^{}, B^\dagger_{\lambda}\}\rangle&=&N\\
\langle[ B_{\lambda}^{}, B^\dagger_{\lambda} ] \rangle&=&N(1-2n_{\rm F}^{})
         \end{array}$
 &
    $\begin{array}{rcl}
     \langle \{B_{\lambda}^{}, B^\dagger_{\lambda} \}\rangle&=&N(1+2n_{\rm B}^{})\\
    \langle [ B_{\lambda}^{},\, B^\dagger_{\lambda}] \rangle&=&N
 \end{array}$\\
    \hline
    \end{tabular}
    }
    \caption{Thermal expectation values of the commutator and anticommutator of the collective bath operators $B_{\lambda}^{}$, $B^\dagger_{\lambda}$ for clusters of $N$ qubits and $N$ oscillators.}
    \label{table:B_excpected_values}
\end{table}

Since the ancillas of a cluster are uncorrelated and each is diagonal in its energy basis, cross terms between different ancillas have zero expectation value, and the thermal averages of the commutator and anticommutator of $B_{\lambda}^{}$ are $N$ times those of a single ancilla,
\begin{equation} \label{eq:GenericConmmutation}
    \begin{split}
    \langle [ B_{\lambda}^{},B^\dagger_{\lambda}]\rangle&=N\left[1+(\lambda-1)n(\lambda)\right],
    \\
    \langle\{B_{\lambda}^{},B^\dagger_{\lambda}\}\rangle&=N\left[1+(\lambda+1)n(\lambda)\right].
    \end{split}
\end{equation}
In particular, $\langle B_{\lambda}^{} B^\dagger_{\lambda} -\lambda \, B^\dagger_{\lambda} B_{\lambda}^{}\rangle=N$; the corresponding operator identity
$B_{\lambda}^{} B^\dagger_{\lambda} -\lambda \, 
B^\dagger_{\lambda} B_{\lambda}^{}=N$, holds exactly for oscillators ($\lambda=1$), while for qubits ($\lambda=-1$) it holds only in expectation value, the operator $\{S_-,S_+\}$ containing cross terms $\sigma_-^{(i)}\sigma_+^{(j)}$, $i\neq j$, that vanish on average. The two cases are summarized in Table~\ref{table:B_excpected_values}. By using the operators $S_\pm$ or $B$ in the generic interaction Eq.~\eqref{eq:Vi-General-Interaction-Hamiltonian} we obtain the master equations~\eqref{eq:GenericMasterEquation} of the main text for either type of bath statistics.

\section{Bloch equations}\label{Apx:Bloch Equations}

\begin{figure}[t]
    \centering
     \begin{tikzpicture}
        \node (img){\includegraphics[width=0.45\textwidth]{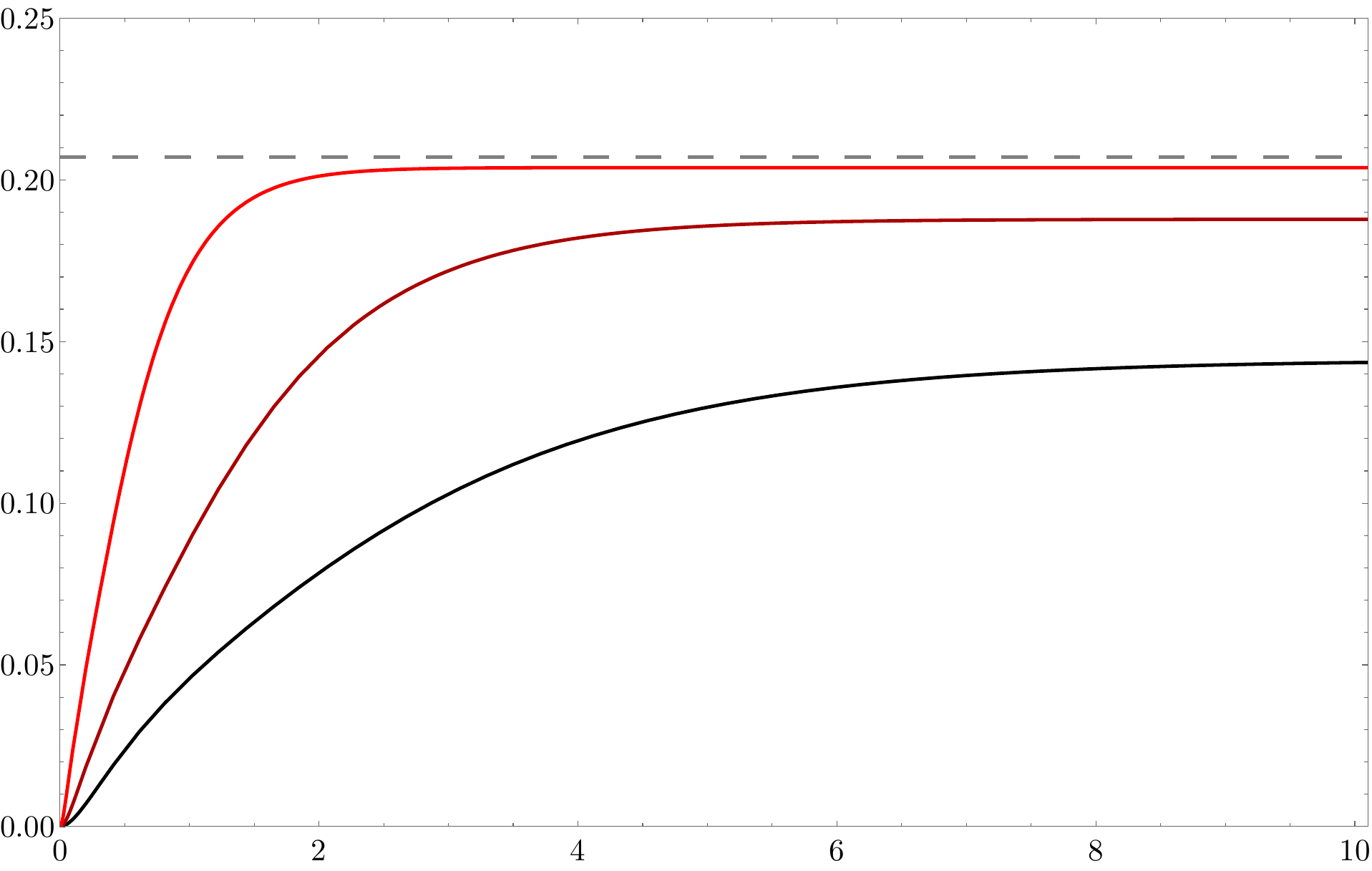}};
        \node[above=of img, node distance=0cm, xshift=1.5cm, yshift=-2.1cm]
        {\textcolor{gray}{\small{\textit{Ergotropy limit ($N\gg 1$)}}}};
        \node[above=of img, node distance=0cm, xshift=1.2cm, yshift=-3.8cm]
        {\textcolor{black}{$N=1$}};
        \node[above=of img, node distance=0cm, xshift=-1.2cm, yshift=-3.2cm]
        {\textcolor{purple}{$N=2$}};
        \node[above=of img, node distance=0cm, xshift=-2.1cm, yshift=-2.6cm]
        {\textcolor{red}{$N=5$}};
        \node[above=of img, node distance=0cm, xshift=0cm, yshift=-6.8cm]
        {$\omega\,t$};
    \end{tikzpicture}
    \caption{Ergotropy of the battery during charging, obtained from the time-dependent solution of the Bloch equations (Appendix~\ref{Apx:Bloch Equations}) for clusters of $N=1,2,5$ ancillas at zero temperature. The gray dashed line is the large-cluster limit $(\sqrt{2}-1)/2$ of Eq.~(\ref{eq:Optimized_Ergotropy}).}
    \label{fig:Ergotropy_1Q-system}
\end{figure}
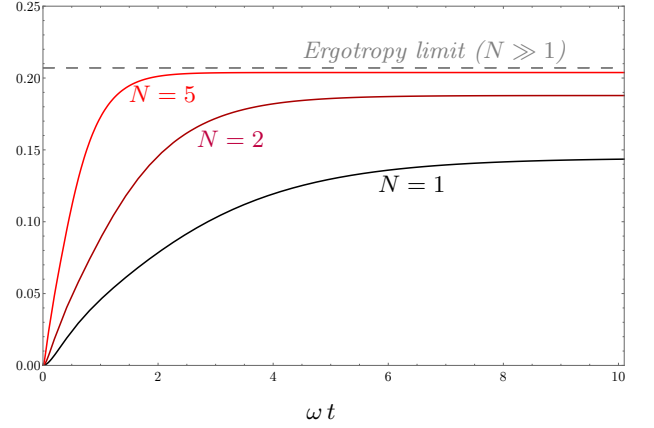

With $\langle\sigma_i\rangle={\rm tr}\{\sigma_i\rho\}$, $i=\left\{x,y,z\right\}$, the master equation \eqref{eq:GenericMasterEquation} yields a closed linear system for the Bloch vector,
\begin{equation}\label{eq:DifferentialBlochEquation}
    \frac{d}{dt}\langle\vec{\sigma}\rangle=\textbf{B}\langle\vec{\sigma}\rangle+\vec{c}\,,
\end{equation}
where

\begin{equation}\label{eq:BlochMatrix}
   \textbf{B}=\begin{pmatrix}
       -\Gamma&-\omega&\Omega\\
       \omega&-\Gamma&0\\
       \Omega&0&-\gamma
   \end{pmatrix}
   \qquad   {\rm and}    \qquad
    \vec{c}=\begin{pmatrix}
    c_x \\
    0 \\
    -c_z
\end{pmatrix} 
\end{equation}
with the rates $\Gamma$, $\gamma$, $\Omega$ and the sources $c_x^{}$, $c_z^{}$ of Table~\ref{table:parametros(B,B)}, in the notation of \cite{R-Ancheyta-Enhanced-2021}. Their explicit $\lambda$ dependence, obtained from Eq.~\eqref{eq:GenericConmmutation}, is given in Table~\ref{table:parameters-Lambda}. Equation~\eqref{eq:DifferentialBlochEquation} is a linear system with constant coefficients; its time-dependent solution is elementary but lengthy and we do not reproduce it. All components relax on a time scale set by $\Gamma$ and $\gamma$ to the steady state derived below. Figure~\ref{fig:Ergotropy_1Q-system} shows the ergotropy evaluated on the transient solution for a battery initially in its ground state, for clusters of $N=1,2,5$ ancillas at zero temperature and with the couplings of Eq.~\eqref{eq:OptimizationIdentity}. Larger clusters charge faster and to a higher value, and at long times $\omega t$ the ergotropy converges to Eq.~\eqref{eq:Ergotropy(lambda)}, approaching $(\sqrt{2}-1)/2$ for $N\gg1$.

\begin{table}[b]
    \centering
{  \renewcommand{\arraystretch}{1.8}
    \begin{tabular}{|c|c|}\hline
    $\begin{array}{rcl}
        \gamma&=&f_2^2\langle \{B_{\lambda}^{},\, B^\dagger_{\lambda}\}\rangle\\
        \Omega&=&f_1^{}\cdot f_2^{}\langle \{B_{\lambda}^{},\, B^\dagger_{\lambda}\}\rangle\\
         \Gamma&=&(2f_1^2+f_2^2/2)\langle \{B_{\lambda}^{},\, B^\dagger_{\lambda}\}\rangle\\
         c_x^{}&=&2(f_1^{}/f_2^{})c_z^{}\\
         c_{z}^{}&=&f_2^2\langle [B_{\lambda}^{},\, B^\dagger_{\lambda} ] \rangle\\  
    \end{array}$\\
         \hline
    \end{tabular}
}
    \caption{Parameters of the Bloch matrix (\ref{eq:BlochMatrix}) in terms of the thermal expectation values of the commutator and anticommutator of the collective bath operators.}
\label{table:parametros(B,B)}
\end{table}

\begin{table}[t]
    \centering
     {  \renewcommand{\arraystretch}{1.8}
    \begin{tabular}{|c|}\hline
    $\begin{array}{rcl}
    \gamma(\lambda)&=&f_2^2N\left[1+(\lambda+1)n(\lambda)\right]\\ 
    \Omega(\lambda)&=&f_1^{}f_2^{}N\left[1+(\lambda+1)n(\lambda)\right]\\
    \Gamma(\lambda)&=&(2f_1^2+f_2^2/2)N\left[1+(\lambda+1)n(\lambda)\right]\\
    c_x(\lambda)&=&2(f_1^{}/f_2^{})c_z\\
    c_z(\lambda)&=&f_2^2N\left[1+(\lambda-1)n(\lambda)\right]\\  
    \end{array}$\\
         \hline
    \end{tabular}
    }
    \caption{$\lambda$-dependent parameters of the Bloch matrix, obtained by inserting Eq.~(\ref{eq:GenericConmmutation}) in Table~\ref{table:parametros(B,B)}.}
    \label{table:parameters-Lambda}
\end{table}

The steady state follows from $d\langle\sigma_i\rangle/dt= {\rm tr}\{\sigma_i\ d\rho/dt\}=0$. The second and third rows of Eq.~\eqref{eq:DifferentialBlochEquation} give $\langle\sigma_y\rangle_{\rm _{SS}}=\omega\langle\sigma_x\rangle_{\rm _{SS}}/\Gamma$ and $\langle\sigma_z\rangle_{\rm _{SS}}=\left(\Omega\langle\sigma_x\rangle_{\rm _{SS}}-c_z\right)/\gamma$, and the first row then yields
\begin{equation}\label{eq:SigmaX_SS_A}
    \langle\sigma_x\rangle_{\rm _{SS}}^{}=\frac{f_1^{}f_2^{}\langle [ B_{\lambda}^{},\, B^\dagger_{\lambda} ] \rangle\,\Gamma}{\Gamma^2+\omega^2-(f_1^{}/f_2^{})^2\,\gamma\,\Gamma}\ .
\end{equation}
Using the temperature dependence of the bath averages (Table~\ref{table:B_excpected_values}), the three components can be written compactly as

\begin{eqnarray}\label{eq:SigmaX_SS_B}
    \langle\sigma_x\rangle_{\rm _{SS}}&=&\frac{f_1^{}f_2^{} \ r(T)}{Ns(T)+\omega^2} ,
    \\
  \langle\sigma_y\rangle_{\rm _{SS}}&=&\frac{\omega f_1^{}f_2^{}\langle [ B_{\lambda}^{},\, B^\dagger_{\lambda} ] \rangle}{Ns(T)+\omega^2},
  \\
%
\langle\sigma_z\rangle_{\rm _{SS}}&=&\frac{f_1^2\,r(T)}{Ns(T)+\omega^2}-\frac{\langle [ B_{\lambda}^{},B^\dagger_{\lambda} ] \rangle}{\langle \{B_{\lambda}^{}, B^\dagger_{\lambda}\}\rangle}\,,
\end{eqnarray}
with
\begin{eqnarray}\label{eq:SigmaX-Y_SS_S}
            r(T)&=&\langle [ B_{\lambda}^{}, B^\dagger_{\lambda} ] \rangle\Gamma
            \\
            &=&\langle [ B_{\lambda}^{},B^\dagger_{\lambda} ] \rangle\langle \{B_{\lambda}^{}, B^\dagger_{\lambda}\}\rangle(2f_1^2+f_2^2/2) 
        \\
        Ns(T)&=&\langle \{B_{\lambda}^{}, B^\dagger_{\lambda}\}\rangle^2(2f_1^2+f_2^2/2)(f_1^2+f_2^2/2).
\end{eqnarray}
Note that the last term of $\langle\sigma_z\rangle_{\rm _{SS}}$ equals $\tanh(\hbar\omega/2k_{\!B}^{}T)$ for both bath statistics [Eq.~\eqref{eq:identity}]. Finally, inserting Eq.~\eqref{eq:GenericConmmutation},

\begin{eqnarray}
  \langle\sigma_x^{\lambda}\rangle_{\rm _{SS}} &=&
  \frac{Nf_1^{}f_2^{}\,(2f_1^2+f_2^2/2)\left(1+2\lambda\,n(\lambda)\right)}{s(T)+\omega^2/N}\,,
    \label{eq:GenericSigmaX}\\
    \langle\sigma_y^{\lambda}\rangle_{\rm _{SS}} &=&
    \frac{\omega f_1^{}f_2^{}\left[1+(\lambda-1)n(\lambda)\right]}{s(T)+\omega^2/N}\,,
    \label{eq:GenericSigmaY}\\
%
   \langle\sigma_z^\lambda\rangle_{_{\rm SS}} &=&
   \frac{f_1}{f_2}\langle\sigma_x^{\lambda}\rangle_{\rm _{SS}}-\tanh\left({\hbar\omega}/{2k_{\!B}^{}T}\right).
   \label{eq:GenericSigmaZ}
\end{eqnarray}
In Eq.~\eqref{eq:GenericSigmaX} we used $[1+(\lambda-1)n(\lambda)][1+(\lambda+1)n(\lambda)]=1+2\lambda n(\lambda)$, valid for $\lambda=\pm1$.

\section{Work rate of the collision model}\label{Apx:Cost}

Between switchings the total Hamiltonian $\mathcal{H}$ is time independent, so all work enters through the switching of $\mathcal{H}_{\!I}^{}=\mathcal{V}_{\!I}^{}/\sqrt{\tau}$ on and off. In the boundary-driven formulation of \cite{R-Ancheyta-Enhanced-2021} the work rate is

\begin{equation}\label{eq:Wdot-def}
  \dot{W}=-\tfrac{1}{2}\big\langle\,[\mathcal{V}_{_I},[\mathcal{V}_{_I},\mathcal{H}_{_S}+\mathcal{H}_{_B}]]\,\big\rangle=+\tfrac{1}{2}\big\langle\,[\mathcal{V}_{_I},[\mathcal{H}_{_0},\mathcal{V}_{_I}]]\,\big\rangle ,
\end{equation}
with $\mathcal{H}_{0}^{}\equiv\mathcal{H}_{S}^{}+\mathcal{H}_{B}^{}$ and the average taken over $\rho(t)\otimes\rho_{_B}^{}$. 

At resonance the orthogonal part of Eq.~\eqref{eq:Vi-General-Interaction-Hamiltonian} commutes with $\mathcal{H}_{_0}$ and only the parallel part contributes to the inner commutator,
\begin{equation}\label{eq:inner}
[\mathcal{H}_{_0},\mathcal{V}_{_I}]=f_1\hbar\omega\,\sigma_z\otimes\big(B_{_\lambda}^{\dagger}-B_{_\lambda}\big).
\end{equation}
Inserting Eq.~\eqref{eq:inner} into Eq.~\eqref{eq:Wdot-def}, using $\sigma_z^2=\mathbb{1}$, and dropping $\langle B_{\lambda}^{2}\rangle=\langle B_{\lambda}^{\dagger2}\rangle=0$ (the ancillas are diagonal in the energy basis) gives Eq.~\eqref{eq:Wdot-general}. The same procedure applied to $\mathcal{H}_{\! S}^{}$ alone gives the internal-energy rate
\begin{equation}\label{eq:Udot}
  \frac{dU}{dt}=\frac{\hbar\omega f_2}{2}
  \big\langle\{B_{_\lambda},B_{_\lambda}^{\dagger}\}\big\rangle
  \left[f_1\langle\sigma_x\rangle
        -f_2\Big(\langle\sigma_z\rangle+\tanh\frac{\hbar\omega}{2k_{_B}T}\Big)\right],
\end{equation}
and the heat rate follows from the first law $dU/dt=\dot{Q}+\dot{W}$. Substituting the steady-state solution \eqref{eq:GenericSigmaZ} into Eq.~\eqref{eq:Udot} gives $dU/dt|_{\rm SS}=0$ identically, so that $\dot{Q}_{\rm SS}=-\dot{W}_{\rm SS}$. In writing Eqs.~\eqref{eq:Wdot-general} and \eqref{eq:Udot} we used

\begin{equation}\label{eq:identity}
  \big\langle[B_{_\lambda},B_{_\lambda}^{\dagger}]\big\rangle
  =\big\langle\{B_{_\lambda},B_{_\lambda}^{\dagger}\}\big\rangle
   \tanh\!\left(\frac{\hbar\omega}{2k_{_B}T}\right),
\end{equation}
which follows from Eq.~\eqref{eq:GenericConmmutation}: with $x\equiv\hbar\omega\beta$, $1+(\lambda-1)n=(e^{x}-1)/(e^{x}-\lambda)$ and $1+(\lambda+1)n=(e^{x}+1)/(e^{x}-\lambda)$, whose ratio is $\tanh(x/2)$ for either $\lambda$. Equation~\eqref{eq:Wdot-largeN} follows from Eq.~\eqref{eq:Wdot-general} with $\langle\sigma_x\rangle_{_{\rm SS}}$ from Sec.~\ref{sec:Steady state for a qubit battery}, using $f_1-f_2\mathcal{C}_0/2=f_1^3/(f_1^2+f_2^2/2)$.

\bibliography{citesArticle}

\end{document}